\documentclass[12pt, reqno]{amsart}
\usepackage{amssymb}
\usepackage{mathrsfs}
\usepackage{mathpazo}
\usepackage[colorlinks=true, breaklinks=true]{hyperref}
\usepackage{bm}
\usepackage[latin1]{inputenc}
\usepackage{graphicx}
\newtheorem{theorem}{Theorem}

\newtheorem{corollary}[theorem]{Corollary}

\newtheorem{definition}[theorem]{Definition}

\newtheorem{lemma}[theorem]{Lemma}

\newtheorem{proposition}[theorem]{Proposition}
\newtheorem{remark}[theorem]{Remark}

\newcommand{\bea}{\begin{eqnarray}}
\newcommand{\eq}{\end{eqnarray}}
\newcommand{\eea}{\end{eqnarray}}
\newcommand{\bqn}{\begin{eqnarray*}}
\newcommand{\beaa}{\begin{eqnarray*}}
\newcommand{\eqn}{\end{eqnarray*}}
\newcommand{\eeaa}{\end{eqnarray*}}
\newcommand{\bpr}{\begin{proposition}}
\newcommand{\epr}{\end{proposition}}

\newcommand{\cal}{\mathcal}

\numberwithin{equation}{section}
\numberwithin{theorem}{section}
\begin{document}
\title{Large deviation principles for stochastic volatility models with reflection and three faces of the Stein and Stein model}
\author{Archil Gulisashvili}\let\thefootnote\relax\footnotetext{Department of Mathematics, Ohio University, Athens OH 45701; e-mail: gulisash@ohio.edu}

\date{}

\begin{abstract}
We introduce stochastic volatility models, in which the volatility is described by a time-dependent nonnegative function of a reflecting diffusion. The idea to use reflecting diffusions as building blocks of the volatility
came into being because of a certain volatility misspecification in the classical Stein and Stein model. A version of this model that uses the reflecting Ornstein-Uhlenbeck process as the volatility process is a special example of a stochastic volatility model with reflection.
The main results obtained in the present paper are sample path and small-noise large deviation principles for the log-price process in a stochastic volatility model with reflection under rather mild restrictions. We use these results to study the asymptotic behavior of binary barrier options and call prices in the small-noise regime.
\end{abstract}

\maketitle

\vspace{0.2in}
\vspace{0.2in}
\\ 
\\
\begin{flushright}
Dedicated to the memory of E. M. Stein.
\end{flushright}
\vspace{0.2in}
\section{Introduction. Three Faces of the Stein and Stein Model}\label{S:I}
In this paper, we introduce stochastic volatility models with reflection and establish sample path and small-noise large deviation principles for the log-price process associated with such a model. We also study small-noise asymptotic behavior of call pricing functions and the implied volatility. 

Sample path large deviation principles go back to the celebrated work of Varadhan (see \cite{V1}) and Freidlin and 
Wentzell (see \cite{FW}). We refer the reader to \cite{DZ,DS,V2} for more information about large deviations. 
Sample path and small-noise large deviation principles have numerous applications in the theory of stochastic volatility models 
(see, e.g., \cite{BC,CMD,FZ,G2,Gul1,Ph,R}). 

The present paper is dedicated to the memory of Elias M. Stein (1931-2018), a prominent mathematician, who for many years had been one of the most influential leaders in the field of harmonic analysis. His mathematical legacy and the impact of his research are well illustrated in 
\cite{AA,F}. In 1991, Elias Stein and Jeremy Stein published the paper \cite{SS}, in which they introduced a stochastic volatility model (the Stein and Stein model) that is currently considered one of the classical stochastic volatility models of financial mathematics. 

The idea to use reflecting diffusions as building blocks of the volatility 
came into being in the present paper because of a certain volatility misspecification in \cite{SS}. More details will be provided below. The volatility process in a model with reflection is described by a time-dependent function of a reflecting diffusion on the half-line.  In this section, we will discuss three versions (faces) of the Stein and Stein model. In the first version, the volatility is modeled by the Ornstein-Uhlenbeck process (this is the original version introduced in \cite{SS}), while in the second and the third version, the volatility process is the absolute value of the Ornstein-Uhlenbeck process and the reflecting Ornstein-Uhlenbeck process, respectively. The third face of the Stein and Stein model is an example of a model with reflection.  

The choice of the arithmetic Ornstein-Uhlenbeck process as the volatility process in \cite{SS} caused certain mostly psychological problems, since a generally accepted paradigm is that the volatility has to be positive. However, according to \cite{Z}, Subsection 3.3.1, and
\cite{LS}, negative volatility in the Stein and Stein model does not cause any conceptual or computational problems. The Stein and Stein model is uncorrelated, which means that Brownian motions driving the asset price process and the volatility process are independent. In such models, marginal distributions of the asset price depend on the integrated variance rather than on the volatility (see, e.g., \cite{G}). Hence, one can use the absolute value of the Ornstein-Uhlenbeck process as the volatility process not changing the model much. However, in \cite{SS}, Stein and Stein claimed the following: ``Before proceeding, we ought to comment on our assumption that volatility is driven by an arithmetic process, which raises the possibility that $\sigma$ can become negative. This formulation is equivalent to putting a reflecting barrier at $\sigma=0$ in the volatility process, since $\sigma$ enters everywhere else in squared fashion" (see \cite{SS}, p. 729). The symbol $\sigma$ in the previous quotation stands for the volatility process in the Stein and Stein model. This misspecification of the volatility, which did not affect the main results in \cite{SS}, was observed by Ball and Roma (see \cite{BR}). They concluded that actually
the absolute value of the Ornstein-Uhlenbeck process, and not the reflecting process, is, in fact, the model for the stochastic volatility in the Stein and Stein model (see \cite{BR}, p. 592). See also a relevant discussion in Subsection 3.3.1 of \cite{Z}.
For the sake of shortness, throughout the rest of the present paper we will write ``the S\&S model" instead of 
``the Stein and Stein model", and also use the abbreviation ``the OU process" instead of ``the Ornstein-Uhlenbeck process". 

Our next goal is to formally introduce the three versions of the S\&S model. We will first comment on the volatility processes associated with these versions. The volatility process $Y^{(1)}$ in the original S\&S model is the OU process. It satisfies the stochastic 
differential equation
\begin{equation}
dY_t^{(1)}=q(m-Y_t^{(1)})dt+\xi dB_t,\quad t\in[0,T],
\label{E:er}
\end{equation}
where $T> 0$ is the time horizon. It is assumed in (\ref{E:er}) that $q\ge 0$, $m\ge 0$, and $\xi> 0$. 
The initial condition for the process $Y^{(1)}$ will be denoted by $y_0$, and it will be assumed that $y_0\in\mathbb{R}$.
The OU process can be represented explicitly as follows:
\begin{equation}
Y_t^{(1)}=e^{-qt}y_0+\left(1-e^{-qt}\right)m+\xi e^{-qt}\int_0^te^{qs}dB_s,\quad t\in[0,T]
\label{E:OUF}
\end{equation}
(see, e.g., \cite{G}, Lemma 1.18). In (\ref{E:er}) and (\ref{E:OUF}), the symbol $B$ stands for a Brownian motion on
a filtered probability space $(\Omega,{\cal F},\{\widetilde{{\cal F}}_t\},\mathbb{P})$, where $\{\widetilde{{\cal F}}_t\}$
is the augmentation of the filtration generated by the process $B$ (see \cite{KaS}, Definition 7.2 in Chapter 2, for the definition of an augmented filtration). The OU process is mean reverting. This is one of the stylistic properties of the volatility. The mean reversion property of the OU process with $q> 0$ can be explained superficially as follows. 
If the process $Y^{(1)}$ diffuses above the long-run mean $m$, then the coefficient
$q(m-Y_t^{(1)})$ in the drift term of (\ref{E:er}) becomes negative, and since the drift term is associated with 
the direction of change of $Y^{(1)}$,
it pulls the process $Y^{(1)}$ towards its mean $m$. A similar observation can be made in the case where $Y^{(1)}$ diffuses below $m$.

Let us set $Y^{(2)}=|Y^{(1)}|$, and let $Y^{(3)}$ be the OU process instantaneously reflected at zero. For the process $Y^{(3)}$, we assume that $y_0\ge 0$. Reflecting diffusions will be discussed in Section \ref{S:22}. The processes $Y^{(k)}$, $1\le k\le 3$, are adapted to the filtration $\{\widetilde{{\cal F}}_t\}$. They are used as the volatility processes in the models that will be introduced next.

Consider the following three stochastic volatility models (the three faces of the S\&S model):
\begin{equation}
dS_t^{(k)}=\mu S_t^{(k)}dt+Y_t^{(k)}S_t^{(k)}d(\overline{\rho}W_t+\rho B_t),\quad t\in[0,T],\quad k=1,2,3.
\label{E:SSmim}
\end{equation}
In the previous models, $\mu\in\mathbb{R}$ is the drift coefficient, while the processes $Y^{(k)}$,
$1\le k\le 3$, describing the stochastic volatility, are such as above.
The process $W$ in (\ref{E:SSmim}) is a Brownian motion defined on the probability space $(\Omega,{\cal F},\mathbb{P})$, 
and it is assumed that the processes $W$ and $B$ are independent. 
We will denote by $\{{\cal F}_t\}$ the augmentation of the filtration generated by the processes $W$ and $B$. 
The processes $S^{(k)}$, $1\le k\le 3$, describing the 
stochastic behavior of the asset price, are adapted to the filtration $\{{\cal F}_t\}$. The number 
$\rho\in(-1,1)$ appearing in (\ref{E:SSmim}) is the correlation parameter, and we use the standard notation 
$\overline{\rho}=\sqrt{1-\rho^2}$. The parameter $\rho$ characterizes the correlation between the process 
$Z=\overline{\rho}W+\rho B$ driving the asset price and the process $B$ driving the volatility. 

We have already mentioned above that the model in (\ref{E:SSmim}) with $k=1$ and $\rho=0$ is the S\&S model introduced in \cite{SS}. 
An important achievement of Stein and Stein was that they found explicit formulas representing the distribution of the underlying and the price of the call option in terms of the Fourier transform. To the author's knowledge, \cite{SS} was the first paper, where Fourier analysis methods were used in the theory of stochastic volatility models. The same model of option pricing as that in the S\&S modle was developed earlier by Scott (see \cite{Sco}, Section 2). However, in \cite{Sco}, Monte Carlo methods were used to estimate option prices, and no analytical formulas were obtained. 
The asymptotic behavior of asset price densities, call prices, and the implied volatility in the S\&S model was studied in
\cite{GS3} (see also \cite{G}). A correlated version ($\rho\neq 0$) of the S\&S model was developed in the paper \cite{SZ} of Schobel and Zhu. 

It is not hard to see that the process $Y^{(1)}$ defined by (\ref{E:OUF}) is a continuous Gaussian
process with the mean function $m(t)=e^{-qt}y_0+\left(1-e^{-qt}\right)m$, $t\in[0,T]$,
and the covariance function $C(t,s)=\xi^2(2q)^{-1}[e^{-q|t-s|}-e^{-q(t+s)}]$, $t,s\in[0,T]$.
The second face of the S\&S model is the model in (\ref{E:SSmim}) with $k=2$. The volatility in this model is the absolute value of a Gaussian process, more precisely, 
$Y^{(2)}_t=|Y_t^{(1)}|$, $t\in[0,T]$. More general stochastic volatility models, in which the volatility follows the absolute value of a Gaussian process, were studied in \cite{GVZ1} and \cite{GVZ2} in the case where $\rho=0$. There are also numerous examples of stochastic volatility models, where the volatility is a nonnegative function of a Volterra type Gaussian process (see, e.g., \cite{FZ,G1,G2,Gul1,CP}). We have chosen only these references here 
because all of them are related to the main subjects of the present paper, which are sample path and small noise large deviation principles
for log-prices in stochastic volatility models. 
 
The third face of the S\&S model is the model in (\ref{E:SSmim}) with $k=3$. The volatility process in this case is the reflecting OU process
$Y^{(3)}$. The third face of the S\&S model is an interesting special example of a stochastic volatility model with reflection. 

The three faces of the S\&S model have many dissimilar features. The transition densities $p_1$ and $p_2$ of the processes 
$Y^{(1)}$ and $Y^{(2)}$, which are known explicitly (see, e.g., \cite{G}, formulas (1.19) and (1.23)), are distinct.
However, as it has been already mentioned, when $\rho=0$, the marginal distributions of the 
asset price processes $S^{(1)}$ and $S^{(2)}$ are identical (see \cite{G} for more details). 
The reflecting OU process $Y^{(3)}$ is a Markov process (see, e.g., Theorem 1.2.2 in \cite{P}).
In \cite{RS}, Ricciardi and Sacerdote showed that the transition density $p_3$ of this process 
is the unique solution to a certain Volterra type integral equation, while in \cite{L}, Linetsky found a spectral representation of $p_3$. 
It is important to mention here that if the long-run mean $m$ of the Ornstein-Uhlenbeck process is equal to zero, then the processes $Y^{(2)}$
and $Y^{(3)}$ are equal in law (see Remark \ref{R:three} below for more information). Therefore, the densities $p_2$ and $p_3$ coincide if $m=0$. This can also be shown by comparing known expressions for $p_2$ 
and $p_3$. As we mentioned above, an explicit representation for the density $p_2$ can be found in \cite{G}.
In \cite{GNR}, the transition function associated with the process $Y^{(3)}$ was characterized (see (4.10) in \cite{GNR}). It is clear that an explicit formula for the density $p_3$ can be obtained by differentiating the functions appearing in the formula for the transition function. The formula described in the previous sentence was rediscovered in \cite{XXY}, Theorem 2.1. It follows from the above-mentioned results 
that if $m=0$, then $p_2=p_3$. 
If $m\neq 0$, then the densities $p_2$ and $p_3$ are different. In \cite{BR}, these densities were compared numerically for a certain set of model parameters such that $m\neq 0$ (see Figure 1 in Appendix A of \cite{BR}). The graphs in Figure 1 show that close to the barrier, the barrier effect is more pronounced for 
$p_2$ than for $p_3$, while far from the barrier, the values of the densities almost coincide. See also Remark \ref{R:rui} in Section
\ref{S:222}.

We will next briefly comment on the structure of the present paper. Section \ref{S:22} of the paper deals with general time-inhomogeneous reflecting diffusions on the half-line. In Section 
\ref{S:222}, stochastic volatility models with reflection are introduced, and the main results obtained in the present paper are formulated and discussed (Theorems \ref{T:27} and \ref{T:17}, and also Corollaries \ref{C:2} and \ref{C:1}). The proof of the general sample path large deviation principle established in Theorem \ref{T:27} is given in Subsection 
\ref{SS:p27}. In the proofs of the results formulated in Section \ref{S:222}, we borrow some ideas from \cite{GGG} and \cite{Gul1}.
The last section of the paper (Section \ref{S:33}) is devoted to
large deviation style asymptotic formulas in the small-noise regime for binary barrier options and call pricing functions. 

\section{Time-Inhomogeneous Reflecting Diffusions}\label{S:22}
This section deals with reflecting diffusions and sample path large deviation principles for them. A good source of information about such diffusions is \cite{GS}, Section 23, and the book \cite{P}. Various large deviation principles for reflecting diffusions were obtained in 
\cite{BC,BZ,CMD,DP,D,HMS,SSS}. Our approach is based on the Skorokhod map, the contraction principle, and a large deviation principle for solutions of multidimensional diffusion equations with predictable coefficients established in \cite{CF} by Chiarini and Fischer. Note that some of the ideas exploited in \cite{CF} were used in \cite{GGG} to study stochastic volatility models generalizing the fractional Heston model.  

Let $a$ and $c$ be jointly continuous functions on $[0,T]\times[0,\infty)$, and let us assume that these functions 
are locally Lipschitz continuous in the second variable, uniformly in time, that is, for every $r> 0$ there exists $L_r> 0$ such that
\begin{equation}
|a(t,x)-a(t,y)|+|c(t,x)-c(t,y)|\le L_r|x-y|,
\label{E:L}
\end{equation}
for all $t\in[0,T]$ and $x,y\in[0,r]$. We also assume that $a$ and $c$ satisfy
the sublinear growth condition in the second variable, uniformly in time:
\begin{equation}
|a(t,x)|+|c(t,x)|\le C(1+|x|),\,\,x\in[0,\infty),\,\,t\in[0,T].
\label{E:C}
\end{equation}
Our goal is to construct a nonnegative diffusion $Y$ having a stochastic differential of the form
$$
dY_t=a(t,Y_t)dt+c(t,Y_t)dB_t,\quad t\in[0,T],
$$
when it lies inside the open half-line $(0,\infty)$, and is instantaneously reflecting when it hits zero. As in the previous chapter, we denote by $\{\widetilde{{\cal F}}_t\}$ the augmentation of the filtration generated by the Brownian motion $B$. The initial condition for the process $Y$ will be denoted by $y_0$, and it will be assumed that $y_0\ge 0$. 

In the pioneering papers \cite{Sk1,Sk2}, Skorokhod suggested to define the instantaneously reflecting process $Y$ as the solution 
to the equation
\begin{align}
&Y_t=y_0+\int_0^ta(s,Y_s)ds+\int_0^tc(s,Y_s)dB_s+l_t,\quad t\in[0,T],
\label{E:skoro}
\end{align}
where $l$ is an auxiliary stochastic process satisfying the following conditions almost surely: 
$l$ is a continuous nondecreasing process with $l_0=0$, and only the zeros of $Y_t$ can be points where $l_t$ increases. 
The latter condition is equivalent to the following: 
$$
l_t=\int_0^t\mathbb{1}_{\{Y_s=0\}}dl_s,\quad t\in[0,T].
$$
Skorohod proved in \cite{Sk1,Sk2} that the equation in (\ref{E:skoro}) with two unknowns $(Y,l)$ has a unique solution. The instantaneously reflecting process $Y$ is a continuous stochastic process, and it follows from (\ref{E:skoro}) that the process $Y$ is a semimartingale. 
Under the restriction $c(t,0)\neq 0$ all $t\in[0,T]$, the following equality holds:
$l_t=\frac{1}{2}L^0_t$, for all $t\in[0,T]$, where $L^0$ is the local time at zero for the process $Y$ 
(see, e.g., \cite{P}, Theorem 1.3.1). The definition of the local time of a semimartingale can be found in \cite{KaS}, Section 3.3.7, see also \cite{RY}, pp. 209-210. 
\begin{remark}\label{R:dopos}
In this remark, we follow Section 23 in \cite{GS} (see also \cite{P}, Exercise 1.3.1). Suppose $a(t,0)=0$ and
 $c(t,0)> 0$, for all $t\in[0,T]$, and extend the functions $a$ and $c$ to the set $[0,T]\times(-\infty,0)$ 
by $a(t,x)=-a(t,-x)$ and $c(t,x)=c(t,-x)$. Denote by $Z$ the solution to the following stochastic differential equation on $\mathbb{R}$:
$$
dZ_t=a(t,Z_t)dt+c(t,Z_t)dB_t,\quad t\in[0,T],
$$ 
with $Z_0=z_0\ge 0$. Then the process $\widetilde{Z}=|Z|$ is the solution to the equation
\begin{equation}
d\widetilde{Z}_t=a(t,\widetilde{Z}_t)dt+c(t,\widetilde{Z}_t)d\widetilde{B}_t+\widetilde{l}_t,\quad t\in[0,T],
\label{E:erts}
\end{equation}
with $\widetilde{Z}_0=z_0$, and a new Brownian motion defined by $\widetilde{B}_t=\int_0^t\mbox{sign}\,Z_sdB_s$, $t\in[0,T]$. Now, using the unique solvability of the equation in (\ref{E:erts}), we see that
under the restrictions on the functions $a$ and $c$ formulated above, the reflecting process $\widetilde{Z}$ and the process 
$|Z|$ have the same laws.
\end{remark}
\begin{remark}\label{R:three}
The reflecting OU process $Y^{(3)}$, which is the volatility process in the third face of the S\&S model corresponds to the case where 
$a(t,x)=q(m-x)$ and $c(t,x)=\xi$, for all $(t,x)\in[0,T]\times\mathbb{R}^{+}$. The process $Y^{(3)}$ is a time-homogeneous nonnegative diffusion.
It follows from Remark \ref{R:dopos} that if $m=0$, then the processes $Y^{(2)}$ and $Y^{(3)}$ are equal in law. This has already been mentioned in the introduction.
\end{remark}

Let $\varepsilon$ be a small-noise parameter. For every $\varepsilon\in[0,1]$, a scaled version of the equation 
in (\ref{E:skoro}) is given by
\begin{align}
&Y_t^{(\varepsilon)}=y_0+\int_0^ta(s,Y_s^{(\varepsilon)})ds+\sqrt{\varepsilon}\int_0^tc(s,Y_s^{(\varepsilon)})dB_s
+l_t^{(\varepsilon)},\quad t\in[0,T].
\label{E:skoros}
\end{align} 

We will next define the Skorokhod map (see \cite{P} for more details).
Let $\mathbb{C}[0,T]$ be the space of continuous functions on $[0,T]$ equipped with the norm
$||f||=\max_{t\in[0,T]}|f(t)|$ for $f\in\mathbb{C}[0,T]$. The Skorokhod map $\Gamma:\mathbb{C}[0,T]\mapsto\mathbb{C}[0,T]$ is given by
\begin{equation}
(\Gamma f)(t)=f(t)-\min_{s\in[0,t]}(f(s)\wedge 0),\quad t\in[0,T].
\label{E:squad}
\end{equation}
The Skorokhod map is a continuous nonlinear mapping from the space $\mathbb{C}[0,T]$ into itself (see, e.g., \cite{P}, Lemma 1.1.1). Actually, $\Gamma$ maps $\mathbb{C}[0,T]$ into $\mathbb{C}^{+}[0,T]$, where the latter symbol stands for the space of all nonegative functions from 
$\mathbb{C}[0,T]$. It
is also true that $(\Gamma(\alpha f))(t)=\alpha(\Gamma f)(t)$, for any $\alpha\ge 0$, $f\in\mathbb{C}[0,T]$, and $t\in[0,T]$. Moreover, it is easy to see that for all $f\in\mathbb{C}[0,T]$ and $t\in[0,T]$, 
\begin{equation}
|(\Gamma f)(t)|\le 2\max_{s\in[0,t]}|f(s)|.
\label{E:inad1}
\end{equation} 
In addition, we have 
$$
|(\Gamma f_1)(t)-(\Gamma f_2)(t)|\le\max_{s\in[0,t]}|f_1(s)-f_2(s)|,
$$
for all $f_1,f_2\in\mathbb{C}[0,T]$ and $t\in[0,T]$.

The Skorokhod map is related to the solution $Y^{(\varepsilon)}$ of the equation in (\ref{E:skoros})
as follows. Denote
$$
U_t^{(\varepsilon)}=y_0+\int_0^ta(s,Y_s^{\varepsilon})ds+\sqrt{\varepsilon}\int_0^tc(s,Y_s^{\varepsilon})dB_s,\quad t\in[0,T].
$$
Then we have
$$
Y_t^{(\varepsilon)}=(\Gamma U^{(\varepsilon)})(t),\quad t\in[0,T],
$$
and moreover for every $\varepsilon\in(0,1]$, the process $t\mapsto U^{(\varepsilon)}_t$ is the solution to the following stochastic differential equation:
\begin{equation}
dU_t^{(\varepsilon)}=a(t,(\Gamma U^{(\varepsilon)})(t))dt+\sqrt{\varepsilon}
c(t,(\Gamma U^{(\varepsilon)})(t))dB_t,\quad t\in[0,T],
\label{E:quad3}
\end{equation}
with $U_0^{(\varepsilon)}=y_0$, $\varepsilon\in(0,1]$ (see \cite{P}, p. 5). Next, using (\ref{E:squad}), we can rewrite (\ref{E:quad3})
as follows:
$$
dU_t^{(\varepsilon)}=a(t,U^{(\varepsilon)}(t)-\min_{s\in[0,t]}(U^{(\varepsilon)}(s)\wedge 0))dt+\sqrt{\varepsilon}
c(t,U^{(\varepsilon)}(t)-\min_{s\in[0,t]}(U^{(\varepsilon)}(s)\wedge 0))dB_t,
$$
for all $t\in[0,T]$.
\begin{remark}\label{R:rtr}
Throughout the paper, we denote by $\mathbb{C}_{\delta}[0,T]$ a subset of the space $\mathbb{C}[0,T]$ consisting of all the functions $f$,
for which $f(0)=\delta$. By $L^2[0,T]$ will be denoted the space of Lebesgue square-integrable over $[0,T]$ functions equipped with the norm
$$
||f||_{L^2[0,T]}=\left\{\int_0^Tf(t)^2dt\right\}^{\frac{1}{2}},\quad f\in L^2[0,T].
$$
The symbol $\mathbb{H}_0^1[0,T]$ will stand for the Cameron-Martin space associated with Brownian motion, that is, the space of all absolutely continuous functions $f$ on $[0,T]$ such that $f(0)=0$ and $\int_0^T\dot{f}(t)^2dt<\infty$. The norm in the space $\mathbb{H}_0^1[0,T]$ is defined by 
$$
||f||_{\mathbb{H}_0^1[0,T]}=\left\{\int_0^T\dot{f}(t)^2dt\right\}^{\frac{1}{2}},\quad f\in\mathbb{H}_0^1[0,T].
$$
By $\mathbb{H}_{\delta}^1[0,T]$ will be denoted the set consisting of all absolutely continuous functions $f$ on $[0,T]$ such that 
$f(0)=\delta$ and $\int_0^T\dot{f}(t)^2dt<\infty$.
\end{remark}

Our next goal is to prove a sample path large deviation principle for the process 
\begin{equation}
T^{(\varepsilon)}=(\sqrt{\varepsilon}W,\sqrt{\varepsilon}{B},U^{(\varepsilon)}),\quad\varepsilon\in(0,1],
\label{E:zm}
\end{equation} 
with state space $\mathbb{C}[0,T]^2\times\mathbb{C}_{y_0}[0,T]$. 
Let us consider the following system of stochastic differential equations:
\begin{equation}
\begin{cases}
dT^{(\varepsilon),1}_t=\sqrt{\varepsilon}dW_t  \\
dT^{(\varepsilon),2}_t=\sqrt{\varepsilon}dB_t  \\
\displaystyle{dT_t^{(\varepsilon),3}=a(t,(\Gamma T^{(\varepsilon),3})(t))dt+\sqrt{\varepsilon}
c(t,(\Gamma T^{(\varepsilon),3})(t))dB_t}.
\end{cases}
\label{E:sati}
\end{equation}

Chiarini and Fisher obtained in \cite{CF} a sample path large deviation principle for solutions of diffusion equations with locally Lipschitz continuous predictable coefficients satisfying a sublinear growth condition (see Theorem 3.1 in \cite{CF}). 
It is not hard to see that the system defined in (\ref{E:sati}) satisfies the conditions in Theorem 3.1 in \cite{CF}. 
Indeed, the coefficients in the third equation in (\ref{E:sati}) have the following form: $a(t,(\Gamma f)(t))$ and $c(t,(\Gamma f)(t))$, for all
$f\in\mathbb{C}[0,T]$ and $t\in[0,T]$. The predictability and the continuity can be established using the definition of the mapping $\Gamma$ and the continuity of $\Gamma$ on the space $\mathbb{C}[0,T]$. The local Lipschitz continuity and the sublinear growth condition follow from the restrictions on the functions $a$ and $c$ formulated in the beginning of this section, and the simple properties of the Skorohod map mentioned above. Therefore, 
Assumptions A1 and A2 on p. 13 of \cite{CF} hold, and hence Theorem 3.1 in \cite{CF} can be applied to the process $T^{\varepsilon}$. We will
next make several remarks and then formulate a sample path LDP for the process $\varepsilon\mapsto T^{\varepsilon}$.

The controlled equation associated with the system above is three-dimensional. It depends on two controls $f_1,f_2\in L^2[0,T]$, and 
has the following form:
\begin{equation}
\begin{cases}
\varphi_1(t)=\int_0^tf_1(s)ds \\
\varphi_2(t)=\int_0^tf_2(s)ds \\
\varphi_3(t)=y_0+\int_0^ta(s,(\Gamma\varphi_3)(s))ds+\int_0^tc(s,(\Gamma\varphi_3)(s))f_2(s)ds
\end{cases}
\label{E:unik}
\end{equation}
(see (2.4) in \cite{CF}). For fixed $f_1,f_2\in L^2[0,T]$, the system in (\ref{E:unik}) has a unique solution. The unique solvability of the third equation in (\ref{E:unik}) was established in \cite{CF} (see the proof of the validity of Assumption H4 on p. 14 of \cite{CF}).
Note that $f_1(t)=\dot{\varphi}_1(t)$ and $f_2(t)=\dot{\varphi}_2(t)$, for all $t\in[0,T]$.

Let $g\in L^2[0,T]$, and let $\varphi_g\in\mathbb{C}_{y_0}[0,T]$ be the unique solution to the equation 
\begin{equation}
\varphi(t)=y_0+\int_0^ta(s,(\Gamma\varphi)(s))ds+\int_0^tc(s,(\Gamma\varphi)(s))g(s)ds.
\label{E:ert}
\end{equation}
(see Assumption H4 in \cite{CF}). Note that in our setting, 
Assumption H4 holds true (see \cite{CF}, Section 3). Actually, it is clear from (\ref{E:ert}) that $\varphi_g\in H^1_{y_0}[0,T]$.
\begin{remark}\label{R:note}
Under the conditions in (\ref{E:L}) and (\ref{E:C}), assumption H5 in \cite{CF} also holds. The latter assumption is as follows: The mapping
$G:L^2[0,T]\mapsto\mathbb{C}_{y_0}[0,T]$ defined by $Gg=\varphi_g$ is a continuous mapping from $V_r$ into the space $\mathbb{C}[0,T]$,
where $V_r\subset L^2[0,T]$ is the closed ball 
of radius $r> 0$ centered at 
the origin equipped with the weak topology. Since such a ball is a compact set in the weak topology of $L^2[0,T]$, 
its image in $\mathbb{C}[0,T]$ under the mapping $G$ is a compact subset of $\mathbb{C}[0,T]$.
\end{remark}

Let us define the following functional on the space $\mathbb{C}[0,T]^3$:
\begin{equation}
I(\varphi_1,\varphi_2,\varphi_3)=\frac{1}{2}\int_0^T\dot{\varphi}_1(t)^2dt+\frac{1}{2}\int_0^T\dot{\varphi}_2(t)^2dt,
\label{E:nez}
\end{equation}
provided that $\varphi_1,\varphi_2\in H_0^1[0,T]$, and $\varphi_3\in\mathbb{C}_{y_0}[0,T]$ is the unique solution to the equation
\begin{equation}
\varphi(t)=y_0+\int_0^ta(s,(\Gamma\varphi)(s))ds+\int_0^tc(s,(\Gamma\varphi)(s))\dot{\varphi}_2(s)ds.
\label{E:ggu}
\end{equation}
Otherwise, we set $I(\varphi_1,\varphi_2,\varphi_3)=\infty$. The equality in (\ref{E:nez}) can be rewritten as follows: For all
$\varphi_1,\varphi_2\in H_0^1[0,T]$,
$$
I(\varphi_1,\varphi_2,G\dot{\varphi}_2)=\frac{1}{2}\int_0^T\dot{\varphi}_1(t)^2dt+\frac{1}{2}\int_0^T\dot{\varphi}_2(t)^2dt,
$$
and $I(\varphi_1,\varphi_2,\varphi_3)=\infty$, otherwise.

Recall that a rate function on a topological space ${\cal X}$ is a lower semi-continuous mapping 
$I:{\cal X}\mapsto[0,\infty]$ such that
for all $y\in[0,\infty)$, the level set $L_y=\{x\in{\cal X}:I(x)\le y\}$ is a closed subset of ${\cal X}$. It is assumed that $I$ is not identically infinite. A rate function $I$ is called a good rate function if for every $y\in[0,\infty)$, the set $L_y$ is a compact 
subset of ${\cal X}$.

The next assertion can be established by applying Theorem 3.1 in \cite{CF}.
\begin{theorem}\label{T:U1}
Suppose that the functions $a$ and $c$ are locally Lipschitz continuous and the sublinear growth condition holds for them.
Then the process $\varepsilon\mapsto T^{\varepsilon}$ defined in (\ref{E:zm}) satisfies the sample path large deviation principle 
with speed $\varepsilon^{-1}$ 
and good rate function $I$ given by (\ref{E:nez}).
The validity of the large deviation principle means that
for every Borel measurable subset ${\cal A}$ of the space $\mathbb{C}[0,T]^3$, the following estimates hold:
\begin{align*}
&-\inf_{(\varphi_1,\varphi_2,\varphi_3)\in{\cal A}^{\circ}}I(\varphi_1,\varphi_2,\varphi_3)
\le\liminf_{\varepsilon\downarrow 0}\varepsilon
\log\mathbb{P}\left(T^{\varepsilon}
\in{\cal A}\right) 
 \\
&\le\limsup_{\varepsilon\downarrow 0}\varepsilon\log\mathbb{P}\left(T^{\varepsilon}\in{\cal A}\right)
\le-\inf_{(\varphi_1,\varphi_2,\varphi_3)\in\bar{{\cal A}}}I(\varphi_1,\varphi_2,\varphi_3).
\end{align*}
The symbols ${\cal A}^{\circ}$ and $\bar{{\cal A}}$ in the previous estimates stand for the interior and the closure of the set 
${\cal A}$, respectively.
\end{theorem}
\begin{remark}\label{R:lp}
Actually, in Theorem 3.1 in \cite{CF} the Laplace principle is established. However, since $I$ is a good rate function, 
the Laplace principle is equivalent to the LDP.
\end{remark}
\begin{remark}\label{R:rrr}
It also follows from the results in \cite{CF} that the sample path large deviation principle holds for the process $\varepsilon
\mapsto U^{(\varepsilon)}$ with speed $\varepsilon^{-1}$ 
and good rate function $J$ given on $\mathbb{C}[0,T]$ by
\begin{equation}
J(\varphi)=\frac{1}{2}\inf_{f\in\mathbb{H}_0^1[0,T]:\varphi=G\dot{f}}\int_0^T\dot{f}(t)^2dt,
\label{E:remm}
\end{equation}
if the equation $\varphi=G\dot{f}$ is solvable for $f$, and $J(\varphi)=\infty$, otherwise.
\end{remark}

It follows from (\ref{E:ggu}) that
$$
\dot{\varphi}_3(t)=a(t,(\Gamma\varphi_3)(t))+c(t,(\Gamma\varphi_3)(t))\dot{\varphi}_2(t),
$$
and hence, if the function $c$ is strictly positive, we have
$$
\dot{\varphi}_2(t)=\frac{\dot{\varphi}_3(t)-a(t,(\Gamma\varphi_3)(t))}{c(t,(\Gamma\varphi_3)(t))}.
$$
Therefore, the following assertion holds true.
\begin{corollary}\label{C:tyu}
Suppose the conditions in Theorem \ref{T:U1} hold. Suppose also that the function $c$ is strictly positive. Then the good rate function
$I$ can be represented as follows: For all $\varphi_1\in\mathbb{H}_0^1[0,T]$ and $\varphi_3\in\mathbb{H}_{y_0}^1[0,T]$,
$$
I(\varphi_1,M\varphi_3,\varphi_3)=\frac{1}{2}\int_0^T\dot{\varphi}_1(t)^2dt+\frac{1}{2}\int_0^TN\varphi_3(s)^2ds,
$$
where
$$
M\varphi_3(t)=\int_0^t\frac{\dot{\varphi}_3(s)-a(s,(\Gamma\varphi_3)(s))}{c(s,(\Gamma\varphi_3)(s))}ds
$$
and
$$
N\varphi_3(t)=\frac{\dot{\varphi}_3(t)-a(t,(\Gamma\varphi_3)(t))}{c(t,(\Gamma\varphi_3)(t))}.
$$
It is also true that $I(\varphi_1,\varphi_2,\varphi_3)=\infty$, otherwise.
\end{corollary}

Our next goal is to prove a large deviation principle for the process 
\begin{equation}
F^{\varepsilon}=(\sqrt{\varepsilon}W,\sqrt{\varepsilon}{B},Y^{(\varepsilon)}),\quad\varepsilon\in(0,1].
\label{E:lpo}
\end{equation}

The next definition will be important in the remaining part of the paper.
\begin{definition}\label{D:hat}
The mapping $f\mapsto\widehat{f}$ of the space $\mathbb{H}_0^1[0,T]$ into the space $\mathbb{C}_{y_0}[0,T]$ is defined by
\begin{equation}
\widehat{f}(t)=(\Gamma(G\dot{f}))(t),\quad t\in[0,T].
\label{E:nashel}
\end{equation}
\end{definition}

In Definition \ref{D:hat}, $\Gamma$ is the Skorokhod map, while $G$ is defined in Remark \ref{R:note}. 

\begin{theorem}\label{T:U2}
Suppose that the functions $a$ and $c$ are locally Lipschitz continuous and the sublinear growth condition holds for them.
Then the process $\varepsilon\mapsto F^{(\varepsilon)}$ defined in (\ref{E:lpo}) satisfies the sample path large deviation principle 
with speed $\varepsilon^{-1}$ 
and good rate function $\widetilde{I}$ given on $\mathbb{C}[0,T]^3$ by 
$$
\widetilde{I}(\psi_1,\psi_2,\widehat{\psi}_2)=\frac{1}{2}\int_0^T\dot{\psi}_1(t)^2dt+\frac{1}{2}\int_0^T\dot{\psi}_2(t)^2dt,
$$
for all $\psi_1,\psi_2\in H_0^1[0,T]$, where the mapping $f\mapsto\widehat{f}$ is defined in (\ref{E:nashel}).
In the rest of the cases, $\widetilde{I}(\psi_1,\psi_2,\psi_3)=\infty$.
\end{theorem}

\it Proof. \rm It is clear that $F^{\varepsilon}=V(U^{\varepsilon})$, where $V:\mathbb{C}[0,T]^3\mapsto\mathbb{C}[0,T]^3$
is defined by 
$$
V(f_1,f_2,f_3)=(f_1,f_2,(\Gamma f_3)).
$$ 
Next, using the continuity of the Skorokhod map on the space $\mathbb{C}[0,T]$, 
Theorem \ref{T:U1}, and the contraction principle, we see that the process $\varepsilon\mapsto F^{(\varepsilon)}$ satisfies the sample 
path LDP with speed $\varepsilon^{-1}$ 
and good rate function $\widetilde{I}$ defined on $\mathbb{C}[0,T]^3$ by 
$$
\widetilde{I}(\psi_1,\psi_2,\psi_3)=\frac{1}{2}\int_0^T\dot{\psi}_1(t)^2dt+\frac{1}{2}\int_0^T\dot{\psi}_2(t)^2dt,
$$
if there exists a function $\varphi\in H_{y_0}^1[0,T]$ such that $\psi_3(t)=(\Gamma\varphi)(t)$, $t\in[0,T]$, and simultaneously
$\varphi=G\psi_2$. We also have $\widetilde{I}(\psi_1,\psi_2,\psi_3)=\infty$, otherwise. It is easy to see that if the function $\varphi$ mentioned above exists, then $\psi_3=(\Gamma(G\psi_2))=\widehat{\psi}_2$.

This completes the proof of Theorem \ref{T:U2}.

The next statement can be obtained from Theorem \ref{T:U2}.
\begin{corollary}\label{C:tyuy}
Suppose the conditions in Theorem \ref{T:U2} hold. Suppose also that the function $c$ is strictly positive. Then the good rate function
$\widetilde{I}$ can be represented as follows. Let $\psi_1\in H_0^1[0,T]$ and $\varphi\in H_{y_0}^1[0,T]$. Then
$$
\widetilde{I}(\psi_1,{\cal M}\varphi,(\Gamma\varphi)-y_0)=\frac{1}{2}\int_0^T\dot{\psi}_1(t)^2dt+\frac{1}{2}\int_0^T{\cal N}
\varphi(s)^2ds,
$$
where
\begin{equation}
{\cal M}\varphi(t)=\int_0^t\frac{\dot{\varphi}(s)-a(s,(\Gamma\varphi)(s))}{c(s,(\Gamma\varphi)(s))}ds
\label{E:calt}
\end{equation}
and
$$
{\cal N}\varphi(t)=\frac{\dot{\varphi}(t)-a(t,(\Gamma\varphi)(t))}{c(t,(\Gamma\varphi)(t))}.
$$
In the rest of the cases, $\widetilde{I}(\psi_1,\psi_2,\psi_3)=\infty$.
\end{corollary}

\it Proof. \rm By taking into account Theorem \ref{T:U2}, we see that it suffices to prove the equality 
\begin{equation}
\widehat{{\cal M}\varphi}=(\Gamma\varphi).
\label{E:by}
\end{equation}
By differentiating the functions in (\ref{E:calt}) with respect to $t$,
we obtain
$$
\dot{\varphi}(t)=a(t,(\Gamma\varphi)(t))+c(t,(\Gamma\varphi)(t))[{\cal M}\varphi]^{\prime}(t),\quad t\in[0,T].
$$
We also have $\varphi(0)=y_0$. The previous equalities mean that $\varphi=G({\cal M}\varphi)$. Therefore
$(\Gamma G)({\cal M}\varphi)=(\Gamma\varphi)$, and it follows that the equality in (\ref{E:by}) holds.

The proof of Corollary \ref{C:tyuy} is thus completed.

\section{Stochastic Volatility Models with Reflection}\label{S:222}
In this section, we introduce stochastic volatility models with reflection, and establish large deviation principles for log-price processes in such models. Let $(\Omega,{\cal F},\mathbb{P})$ be a probability space carrying two independent standard Brownian motions $W$ and $B$,
and let $Y$ be a time-inhomogeneous reflecting diffusion satisfying
the equation in (\ref{E:skoro}). It will be assumed that the conditions in (\ref{E:L}) and (\ref{E:C}) hold for the coefficients $a$ and $c$.
Consider a stochastic volatility model, in which the asset price process 
$S_t$, $t\in[0,T]$, satisfies the following stochastic differential equation:
\begin{equation}
dS_t=S_tb(t,Y_t)dt+S_t\sigma(t,Y_t)(\bar{\rho}dW_t+\rho dB_t),\quad S_0=s_0> 0,\quad 0\le t\le T.
\label{E:mood}
\end{equation}
In (\ref{E:mood}), $s_0$ is the initial price, $T> 0$ is the time horizon, $\rho\in(-1,1)$ is the correlation coefficient, and  
$\bar{\rho}=\sqrt{1-\rho^2}$. The functions $b$ and $\sigma$ are continuous functions on $[0,T]\times\mathbb{R}$. 
The equation in (\ref{E:mood}) is considered on a filtered probability space 
$(\Omega,\mathcal{F},\{\mathcal{F}_t\}_{0\le t\le T},\mathbb{P})$, where $\{\mathcal{F}_t\}_{0\le t\le T}$ is the augmentation of the filtration generated by the processes $W$ and $B$.
We will also use the augmentation of the filtration generated by the process $B$, and denote it by
$\{\widetilde{\mathcal{F}}_t\}_{0\le t\le T}$. It is clear that the process $Y$ is adapted to the filtration $\{\widetilde{\mathcal{F}}_t\}_{0\le t\le T}$.

It will be explained next what restriction we impose on the functions $b$ and $\sigma$ appearing in (\ref{E:mood}). This restriction is 
rather mild. The following definitions will be needed.
\begin{definition}\label{D:modc}
A locally bounded function $\omega:[0,\infty)\mapsto[0,\infty)$ is called a modulus of continuity
on $[0,\infty)$, if $\omega(0)=0$ and $\displaystyle{\lim_{u\rightarrow 0}\omega(u)=0}$.
\end{definition}
\begin{definition}\label{D:modco}
Let $\omega$ be a modulus of continuity on $[0,\infty)$.
A function $\lambda$ defined on $[0,T]\times\mathbb{R}$ is called locally $\omega$-continuous,
if for every $\delta> 0$ there exists a number $L(\delta)> 0$ such that for all $x,y\in\overline{B(\delta)}$,
the following inequality holds:
\begin{equation}
|\lambda(x)-\lambda(y)|\le L(\delta)\omega(||x-y||).
\label{E:delta}
\end{equation}
In (\ref{E:delta}), the symbol $||\cdot||$ stands for the Euclidean norm on $[0,T]\times\mathbb{R}$, and $\overline{B(\delta)}$
denotes the closed ball in the space $[0,T]\times\mathbb{R}$ centered at $(0,0)$ and of radius $\delta$.
\end{definition}
We will next formulate the restriction that we impose on the functions $b$ and $\omega$. \\
\\
\it Assumption C. \rm The functions $b$ and $\sigma$ are locally $\omega$-continuous on $[0,T]\times\mathbb{R}$ with respect to some modulus of continuity $\omega$. Moreover, 
the function $\sigma$ is nonnegative and not identically zero on $[0,T]\times\mathbb{R}$. \\
\\ 
If all the conditions formulated above hold, we call the model described by the equation in (\ref{E:mood}) 
a stochastic volatility model with reflection. 
\begin{remark}\label{R:noot}
The third face of the S\&S model is an example of a model with reflection. For this model, the process $Y$ in (\ref{E:mood}) is the reflecting OU process
$Y^{(3)}$. Here we have $b(t,x)=\mu$ for all $(t,x)\in[0,T]\times\mathbb{R}$, while $\sigma(t,x)=x$ for all $(t,x)\in[0,T]\times[0,\infty)$, 
and $\sigma(t,x)=0$
for all $(t,x)\in[0,T]\times(-\infty,0)$ (compare the model in (\ref{E:SSmim}) with $k=3$ and the model in (\ref{E:mood})).
In addition, $a(t,x)=q(m-x)$ and $c(t,x)=\xi$, for all $(t,x)\in[0,T]\times\mathbb{R}$ (see Remark \ref{R:three}). If $\mu=r$, then the 
third face of the S\&S model is a risk-neutral model (see Remark \ref{R:88}).
\end{remark}

The unique solution to the equation in (\ref{E:mood}) is the Dol\'{e}ans-Dade exponential
$$
S_t=s_0\exp\left\{\int_0^tb(s,Y_s)ds-\frac{1}{2}\int_0^t\sigma^2(s,Y_s)ds+\int_0^t\sigma(s,Y_s)
(\bar{\rho}dW_s+\rho dB_s)\right\},\,0\le t\le T,
$$
(see, e.g., \cite{RY}). Therefore, the log-price process $X_t=\log S_t$ satisfies
$$
X_t=x_0+\int_0^tb(s,Y_s)ds-\frac{1}{2}\int_0^t\sigma^2(s,Y_s)ds+\int_0^t\sigma(s,Y_s)
(\bar{\rho}dW_s+\rho dB_s),\,0\le t\le T,
$$
where $x_0=\log s_0$. 

We will work with the following scaled version of the model in (\ref{E:mood}):
$$
dS^{(\varepsilon)}_t=S^{(\varepsilon)}_tb(t,Y_t^{(\varepsilon)})dt
+\sqrt{\varepsilon}S^{(\varepsilon)}_t\sigma\left(t,Y_t^{(\varepsilon)}\right)(\bar{\rho}dW_t+\rho dB_t),
$$
where $0\le t\le T$, and $Y^{(\varepsilon)}$ is the process satisfying the equation in (\ref{E:skoros}). 
The asset price process in the scaled model is given by;
\begin{align}
&S_t^{(\varepsilon)}=s_0\exp\left\{\int_0^tb(s,Y_s^{(\varepsilon)})ds
-\frac{1}{2}\varepsilon\int_0^t\sigma(s,Y_s^{(\varepsilon)})^2ds
+\sqrt{\varepsilon}
\int_0^t\sigma(s,Y_s^{(\varepsilon)})(\bar{\rho}dW_s+\rho dB_s)\right\},
\label{E:frum}
\end{align}
where $0\le t\le T$, while the log-price process is as follows:
\begin{align}
&X^{(\varepsilon)}_t=x_0+\int_0^tb(s,Y_s^{(\varepsilon)})ds
-\frac{1}{2}\varepsilon\int_0^t\sigma(s,Y_s^{(\varepsilon)})^2ds+\sqrt{\varepsilon}
\int_0^t\sigma(s,Y_s^{(\varepsilon)})(\bar{\rho}dW_s+\rho dB_s),
\label{E:frrom} 
\end{align}
where $0\le t\le T$. 

Our next goal is to formulate and prove large deviation principles for the process $\varepsilon\mapsto X^{(\varepsilon)}-x_0$. 
Analyzing the representation for the process $X^{(\varepsilon)}$ given in (\ref{E:frrom}), we see why it was important to establish an LDP for the process 
$F^{\varepsilon}$ defined in (\ref{E:lpo}). It is clear that the components of the process $F^{\varepsilon}$ are building blocks of the process
$X^{(\varepsilon)}$, and our aim is to use the extended contraction principle (see \cite{DZ}) to establish large deviation principles for 
the process $\mapsto X^{(\varepsilon)}-x_0$. Some of the techniques used in such proofs were
developed in \cite{FZ,G1,G2,Gul1,CP} in the case, where the volatility is modeled by a function of a Gaussian process, and in \cite{GGG} for certain non-Gaussian models. In this section, we borrow some ideas employed in the proof of the sample path LDP in Theorem 4.2 of 
\cite{Gul1} (see Subsection 5.6 of \cite{Gul1}). However, there are also significant differences between the two proofs, because the mappings
$f\mapsto\widehat{f}$ used in \cite{Gul1} and in the present paper are very different. Recall that in this paper, 
$\widehat{f}(t)=(\Gamma(G\dot{f}))(t)$, $t\in[0,T]$, where $\Gamma$ is the Skorokhod map, and $G$ is defined in Remark \ref{R:note}, while in \cite{Gul1},
$\widehat{f}(t)=\int_0^tK(t,s)\dot{f}(s)ds$, $t\in[0,T]$,
where $K$ is a Volterra type kernel that is Lebesgue square integrable over $[0,T]^2$. For the sake of convenience, we have decided to steal the notation $\widehat{f}$ from \cite{Gul1}, since certain parts of the proofs in Subsection 5.6 of \cite{Gul1} and in the present section do not depend on a special structure of the mapping $f\mapsto\widehat{f}$.
 
We will next formulate several theorems. They resemble the LDPs obtained in \cite{Gul1}. First, we introduce some notation.
Consider a measurable functional 
$$
\Phi:\mathbb{C}_0[0,T]^2\times\mathbb{C}[0,T]\mapsto\mathbb{C}_0[0,T]
$$ 
defined as follows: For 
$l,f\in\mathbb{H}^1_0[0,T]$ and $h=\widehat{f}\in\mathbb{C}_{y_0}[0,T]$,
\begin{equation}
\Phi(l,f,h)(t)=\int_0^tb(s,\widehat{f}(s))ds+\bar{\rho}\int_0^t\sigma(s,\widehat{f}(s))\dot{l}(s)ds
+\rho\int_0^t\sigma(s,\widehat{f}(s))\dot{f}(s)ds,
\label{E:ref1}
\end{equation}
where $0\le t\le T$. For all the remaining triples $(l,f,h)$, we set $\Phi(l,f,h)(t)=0$, $t\in[0,T]$. 

Let $g\in\mathbb{C}_0[0,T]$, and define
\begin{align*}
&\widetilde{Q}_T(g)
=\inf_{l,f\in\mathbb{H}_0^1[0,T]}\left[\frac{1}{2}\left(\int_0^T\dot{l}(s)^2ds
+\int_0^T\dot{f}(s)^2ds\right):\Phi(l,f,\widehat{f})(t)=g(t),\,
t\in[0,T]\right],
\end{align*}
if the equation appearing on the right-hand side of the previous formula is solvable for $l$ and $f$. If there is no solution, then we set 
$\widetilde{Q}_T(g)=\infty$. It is not hard to see that if the equation $\Phi(l,f,\widehat{f})(t)=g(t)$ is solvable, then 
$g\in\mathbb{H}_0^1[0,T]$.

The next two assertions contain sample path large deviation principles for the log-price process in a 
time-inhomogeneous stochastic volatility model with reflection. At the first glance, these assertions look exactly as the large deviation principles formulated in Theorems 4.2 and 4.3 in \cite{Gul1}. However, there is a significant difference between the LDPs obtained in 
\cite{Gul1} and in the present paper. This difference arises because of the contrasting forms of the mapping 
$f\mapsto\widehat{f}$ in \cite{Gul1} and in this paper.
\begin{theorem}\label{T:27}
Suppose the functions $a$ and $c$ are locally Lipschitz continuous and the sublinear growth condition holds for them. 
Suppose also that Assumption C holds for the functions $b$ and $\sigma$.
Then the process $\varepsilon\mapsto X^{(\varepsilon)}-x_0$ with state space $\mathbb{C}_0[0,T]$ satisfies the sample path large deviation principle with speed $\varepsilon^{-1}$ 
and good rate function $\widetilde{Q}_T$.
The validity of the large deviation principle means that
for every Borel measurable subset ${\cal A}$ of $\mathbb{C}_0[0,T]$, the following estimates hold:
\begin{align*}
&-\inf_{g\in{\cal A}^{\circ}}\widetilde{Q}_T(g)\le\liminf_{\varepsilon\downarrow 0}\varepsilon\log\mathbb{P}\left(
X^{(\varepsilon)}-x_0\in{\cal A}\right) 
 \\
&\le\limsup_{\varepsilon\downarrow 0}\varepsilon\log\mathbb{P}\left(X^{(\varepsilon)}-x_0\in{\cal A}\right)
\le-\inf_{g\in\bar{{\cal A}}}\widetilde{Q}_T(g).
\end{align*}
The symbols ${\cal A}^{\circ}$ and $\bar{{\cal A}}$ in the previous estimates stand for the interior and the closure of the set 
${\cal A}$, respectively.
\end{theorem}
\begin{corollary}\label{C:2}
Suppose the conditions in Theorem \ref{T:27} hold. Suppose also that the volatility function $\sigma$ is strictly positive on 
$[0,T]\times\mathbb{R}$. Then, for all $g\in\mathbb{H}_0^1[0,T]$,
\begin{align}
&\widetilde{Q}_T(g) 
=\inf_{f\in\mathbb{H}_0^1[0,T]}\left[\frac{1}{2}\int_0^T\left[\frac{\dot{g}(s)-b(s,\widehat{f}(s))-\rho\sigma(s,\widehat{f}(s))\dot{f}(s)}
{\bar{\rho}\sigma(s,\widehat{f}(s))}\right]^2ds
+\frac{1}{2}\int_0^T\dot{f}(s)^2ds\right].
\label{E:babe}
\end{align}
\end{corollary}
\begin{remark}\label{R:vbn1}
Under the conditions in Corollary \ref{C:2}, the function $\widetilde{Q}_T:\mathbb{H}_0^1[0,T]\mapsto\mathbb{R}$ is continuous. Indeed,
since $\widetilde{Q}_T$ is a rate function on $\mathbb{C}_0[0,T]$, it is lower semicontinuous on that space. It follows that 
$\widetilde{Q}_T$ is also lower semicontinuous on the space $\mathbb{H}_0^1[0,T]$, since the latter space is continuously embedded into the space $\mathbb{C}_0[0,T]$. The upper semicontinuity of the function $\widetilde{Q}_T$ on the space $\mathbb{H}_0^1[0,T]$ follows from the
fact that this function can be represented as the infimum of a family of functions, which are continuous 
on the space $\mathbb{H}_0^1[0,T]$ (see (\ref{E:babe})). The continuity of those functions on $\mathbb{H}_0^1[0,T]$ can be established as in 
Lemma 6.2 in \cite{Gul1}.
\end{remark}

We will next show how to derive Corollary \ref{C:2} from Theorem \ref{T:27}. Suppose the conditions in Corollary \ref{C:2} hold,
and let $f,g\in\mathbb{H}_0^1[0,T]$. Then, the equation 
\begin{equation}
\Phi(l,f,\widehat{f})(t)=g(t),\quad t\in[0,T],
\label{E:diffr}
\end{equation}
is solvable for $l\in\mathbb{H}_0^1[0,T]$. Moreover, for any such solution, we have 
$$
\dot{l}(s)=\frac{\dot{g}(s)-b(s,\widehat{f}(s))-\rho\sigma(s,\widehat{f}(s))\dot{f}(s)}
{\bar{\rho}\sigma(s,\widehat{f}(s))},\quad s\in[0,T].
$$
The previous statement can be established by differentiating the functions in (\ref{E:diffr}) with respect to $t$, and solving 
the resulting equation for 
$\dot{l}$. Now, it is clear how to finish the proof of Corollary \ref{C:2}. 

Our next goal is to formulate small-noise large deviation principles for the log-price process $\varepsilon\mapsto X^{(\varepsilon)}_T-x_0$
with state space $\mathbb{R}$,
in a time-inhomogeneous stochastic volatility model with reflection.

Let $y\in\mathbb{R}$, $f\in\mathbb{H}_0^1[0,T]$, and put
\begin{align*}
&\Psi(y,f,\widehat{f})=\int_0^T[b(s,\widehat{f}(s))+\rho\sigma(s,\widehat{f}(s))\dot{f}(s)]ds+\bar{\rho}
\left\{\int_0^T\sigma(s,\widehat{f}(s))^2ds\right\}^{\frac{1}{2}}y.
\end{align*}
Define a function on $\mathbb{R}$ as follows:
\begin{align}
&\widetilde{I}_T(x)=\inf_{y\in\mathbb{R},f\in\mathbb{H}_0^1[0,T]}\left[\frac{1}{2}\left(y^2
+\int_0^T\dot{f}(s)^2ds\right):\Psi(y,f,\widehat{f})=x\right],
\label{E:vunzics}
\end{align}
if the equation $\Psi(y,f,\widehat{f})=x$ is solvable, and $\widetilde{I}_T(x)=\infty$, otherwise.
\begin{theorem}\label{T:17}
Suppose the functions $a$ and $c$ are locally Lipschitz continuous and the sublinear growth condition holds for them. 
Suppose also that Assumption C holds for the functions $b$ and $\sigma$. Then the process $\varepsilon\mapsto X^{(\varepsilon)}_T-x_0$ satisfies the small-noise large deviation principle 
with speed $\varepsilon^{-1}$ 
and good rate function $\widetilde{I}_T$ given by (\ref{E:vunzics}). The validity of the large deviation principle means that
for every Borel measurable subset $A$ of $\mathbb{R}$, the following estimates hold:
\begin{align*}
&-\inf_{x\in A^{\circ}}\widetilde{I}_T(x)\le\liminf_{\varepsilon\downarrow 0}\varepsilon\log\mathbb{P}\left(
X_T^{(\varepsilon)}-x_0\in A\right) 
 \\
&\le\limsup_{\varepsilon\downarrow 0}\varepsilon\log\mathbb{P}\left(X_T^{(\varepsilon)}-x_0\in A\right)
\le-\inf_{x\in\bar{A}}\widetilde{I}_T(x).
\end{align*}
The symbols $A^{\circ}$ and $\bar{A}$ in the previous estimates stand for the interior and the closure of the set $A$, respectively.
\end{theorem}

Theorem \ref{T:17} follows from Theorem \ref{T:27}. The previous statement can be established by using the same reasoning as in the derivation of Theorem 4.11 from Theorem 4.2 in Section 4 of \cite{Gul1}. Note that the proof in \cite{Gul1} does not depend on a special form of the mapping $f\mapsto\widehat{f}$.
\begin{remark}\label{R:remis}
A set ${\cal A}\subset\mathbb{C}_0[0,T]$ is called a set of continuity for the rate function $\widetilde{Q}_T$ (see Theorem \ref{T:27}), 
if
$$
\inf_{g\in{\cal A}^{\circ}}\widetilde{Q}_T(g)=\inf_{g\in\bar{{\cal A}}}\widetilde{Q}_T(g).
$$
For such a set, Theorem \ref{T:27} implies that
\begin{equation}
\lim_{\varepsilon\downarrow 0}\varepsilon\log\mathbb{P}\left(X^{(\varepsilon)}-x_0\in{\cal A}\right)
=-\inf_{g\in{\cal A}}\widetilde{Q}_T(g).
\label{E:00}
\end{equation}
A similar definition of a set of continuity can be given for the rate function $\widetilde{I}_T$ in Theorem \ref{T:17},
and an equality similar to that in (\ref{E:00}) can be established.
\end{remark}
\begin{corollary}\label{C:1}
Suppose the conditions in Theorem \ref{T:17} hold. Suppose also that the volatility function $\sigma$ is strictly positive on 
$[0,T]\times\mathbb{R}$. Then, for every $x\in\mathbb{R}$,
\begin{equation}
\widetilde{I}_T(x)=\inf_{f\in\mathbb{H}_0^1[0,T]}\left[\frac{\left(x-\int_0^T[b(s,\widehat{f}(s))+\rho\sigma(s,\widehat{f}(s))\dot{f}(s)]ds\right)^2}
{2\bar{\rho}^2\int_0^T\sigma(s,\widehat{f}(s))^2ds}+\frac{1}{2}\int_0^T\dot{f}(s)^2ds\right].
\label{E:vunz}
\end{equation}
\end{corollary}

Corollary \ref{C:1} can be obtained from Theorem \ref{T:17} as follows. Let $x\in\mathbb{R}$ and $f\in\mathbb{H}_0^1[0,T]$. Then, under the conditions in Corollary \ref{C:1}, the
equation $\Psi(y,f,\widehat{f})=x$ can be solved for $y\in\mathbb{R}$, and for every such solution we have
$$
y^2=\frac{\left(x-\int_0^T[b(s,\widehat{f}(s))+\rho\sigma(s,\widehat{f}(s))\dot{f}(s)]ds\right)^2}
{2\bar{\rho}^2\int_0^T\sigma(s,\widehat{f}(s))^2ds}.
$$
Now, it is clear that Corollary \ref{C:1} holds.
\begin{remark}\label{R:vbn2}
The good rate function function $\widetilde{I}_T$ given by (\ref{E:vunz}) is continuous on $\mathbb{R}$. The proof is similar to that in
Remark \ref{R:vbn1}.
\end{remark}
\begin{remark}\label{R:davito}
It is not hard to see that Corollary \ref{C:1} also holds if for every $f\in\mathbb{H}_0^1[0,T]$, 
\begin{equation}
\int_0^T\sigma(s,\widehat{f}(s))^2ds\neq 0.
\label{E:dev}
\end{equation}
By the continuity of the functions in (\ref{E:dev}), the equality in (\ref{E:dev}) is equivalent to the following condition: 
For every $f\in\mathbb{H}_0^1[0,T]$, there exists $s\in[0,T]$ such that
$\sigma(s,\widehat{f}(s))\neq 0$. The point $s$ in the previous sentence may depend on $f$.
\end{remark}

We will next analyze the condition in (\ref{E:dev}). 
\begin{lemma}\label{L:o}
The following are true: \\
(i)\, Suppose $y_0> 0$. Suppose also that $\sigma(0,y_0)\neq 0$. Then, for every $f\in\mathbb{H}_0^1[0,T]$, 
the condition in (\ref{E:dev}) holds.
\\
(ii)\,Let $y_0=0$, and suppose the functions $a$ and $c$ appearing in the model for the reflecting volatility process 
are such that the function $g$ defined by $g(t)=-\frac{a(t,0)}{c(t,0)}$, $t\in[0,T]$, is Lebesgue square integrable over $[0,T]$. 
Suppose also that $\sigma(s,0)=0$, for all $s\in[0,T]$. Then, the function $f$ given by $f(t)=\int_0^tg(u)du$, $t\in[0,T]$, 
is such that $f\in\mathbb{H}_0^1[0,T]$, and moreover $\sigma(s,\widehat{f}(s))=0$ for all $s\in[0,T]$.
\end{lemma}

\it Proof. \rm Let the conditions in part (i) of Lemma \ref{L:o} hold. We will reason by contradiction. Suppose for some 
$f\in\mathbb{H}_0^1[0,T]$, we have 
$\widehat{f}(t)=0$ for all $t\in[0,T]$. Since $\widehat{f}=\Gamma(G\dot{f})$, and $G\dot{f}(0)=y_0$, we have $\widehat{f}(0)=y_0> 0$.
The previous formula contradicts our original assumption.
This establishes part (i) of Lemma \ref{L:o}.

Suppose the conditions in part (ii) of Lemma \ref{L:o} hold. Then we have $\dot{f}=g\in L^2[0,T]$. It follows that
\begin{equation}
0=\int_0^ta(s,(\Gamma 0)(s))ds+\int_0^tc(s,(\Gamma 0)(s))\dot{f}(s)ds,\quad t\in[0,T].
\label{E:p}
\end{equation} 
Since for every function $g\in L^2[0,T]$, the equation in (\ref{E:ert}) is uniquely solvable, and we denoted its unique solution by $Gg$ 
(see Remark \ref{R:note}), the following equality can be derived from (\ref{E:p}): $G\dot{f}(t)=0$, $t\in[0,T]$. 
Recall that $\widehat{f}=\Gamma (G\dot{f})$ (see (\ref{E:nashel})). It follows that $\widehat{f}(s)=0$, for all $s\in[0,T]$. Now
part (ii) of Lemma \ref{L:o} follows from the condition $\sigma(s,0)=0$, $s\in[0,T]$.

This completes the proof of Lemma \ref{L:o}.
\begin{corollary}\label{C:try}
The following is true for the third face of the Stein and Stein model, that is, the model in (\ref{E:SSmim}) with $k=3$.
Suppose $y_0> 0$. Then,
for every $f\in\mathbb{H}_0^1[0,T]$, the function $\widehat{f}$ is given by $\widehat{f}=\Gamma\varphi_f$, where 
$\varphi_f\in\mathbb{H}_{y_0}^1[0,T]$ is the unique solution to the equation
\begin{equation}
\varphi_f(t)=y_0+qmt-q\int_0^t\Gamma\varphi_f(s)ds+\xi f(t),\quad t\in[0,T].
\label{E:lpos}
\end{equation}
Moreover, the large deviation principle in Theorem \ref{T:17} holds with the rate function $\widetilde{I}_T$ given by 
$$
\widetilde{I}_T(x)=\inf_{f\in\mathbb{H}_0^1[0,T]}\left[\frac{\left(x-\mu T-\rho\int_0^T(\Gamma\varphi_f)(s)\dot{f}(s)ds\right)^2}
{2\bar{\rho}^2\int_0^T(\Gamma\varphi_f)(s)^2ds}+\frac{1}{2}\int_0^T\dot{f}(s)^2ds\right].
$$
\end{corollary}

\it Proof. \rm Corollary \ref{C:try} follows from part (i) of Lemma \ref{L:o}, Remark \ref{R:davito}, and Corollary \ref{C:1}.
We also take into account that for the third face of the S\&S model,
$a(t,u)=q(m-u)$, $c(t,u)=\xi$, and $b(t,u)=\mu$, for all $(t,u)\in[0,T]\times\mathbb{R}^{+}$. Moreover, we can assume that
$\sigma(t,u)=0$ on $[0,T]\times(-\infty,0)$ and
$\sigma(t,u)=u$ on $[0,T]\times[0,\infty)$. The equation in (\ref{E:lpos}) can be obtained from (\ref{E:ert}) with $g=\dot{f}$, Remark
\ref{R:note}, and (\ref{E:nashel}).

The proof of Corollary \ref{C:try} is thus completed.

The case where $y_0=0$ is more complicated. In this case, the set
$$
L_1=\{f\in\mathbb{H}_0^1[0,T]:\sigma(s,\widehat{f}(s))=0\,\,\mbox{for all}\,\,s\in[0,T]\}
$$
is not empty (see part (ii) of Lemma \ref{L:o}). For Gaussian stochastic volatility models, a similar problem was encountered
in \cite{Gul1} (see Lemma 4.10 in \cite{Gul1}). For the third version of the S\&S model, we have
\begin{equation}
L_1=\{f\in\mathbb{H}_0^1[0,T]:(\Gamma\varphi_f)(s)=0\,\,\mbox{for all}\,\,s\in[0,T]\},
\label{E:poy}
\end{equation}
where the function $\varphi_f$ can be determined from (\ref{E:lpos}). We also set $L_2=\mathbb{H}_0^1[0,T]\backslash L_1$.

The following assertion holds in the case where $y_0=0$. The proof is similar to that of Lemma 4.10 in \cite{Gul1}.
\begin{corollary}\label{C:tryes}
Suppose $y_0=0$ in Corollary \ref{C:try}. Then,
for every $f\in\mathbb{H}_0^1[0,T]$, the function $\widehat{f}$ is given by $\widehat{f}=\Gamma\varphi_f$, where 
$\varphi_f\in\mathbb{H}_0^1[0,T]$ is the unique solution to the equation
$$
\varphi_f(t)=qmt-q\int_0^t\Gamma\varphi_f(s)ds+\xi f(t),\quad t\in[0,T].
$$
Moreover, the large deviation principle in Theorem \ref{T:17} holds with the rate function $\widetilde{I}_T$ given by 
$$
\widetilde{I}_T(\mu T)=\frac{1}{2}\min\left\{\inf_{f\in L_1}\int_0^T\dot{f}(t)^2dt,
\inf_{f\in L_2}\left[\frac{\rho^2\left(\int_0^T(\Gamma\varphi_f)(s)\dot{f}(s)ds\right)^2}
{\bar{\rho}^2\int_0^T(\Gamma\varphi_f)(s)^2ds}+\int_0^T\dot{f}(s)^2ds\right]\right\}
$$
and
$$
\widetilde{I}_T(x)=\inf_{f\in L_2}\left[\frac{\left(x-\mu T-\rho\int_0^T\widehat{f}(s)\dot{f}(s)ds\right)^2}
{2\bar{\rho}^2\int_0^T\widehat{f}(s)^2ds}+\frac{1}{2}\int_0^T\dot{f}(s)^2ds\right],
$$
for $x\neq\mu T$.
\end{corollary}

It may be difficult to find a simple explicit 
description of the set $L_1$. We will do it below for a special model that will be introduced next.

Brownian motion with drift is defined by
\begin{equation}
Y_t^{(1)}=at+\xi B_t,\quad t\in[0,T],
\label{E:bm}
\end{equation} 
where $a\ge 0$ and $\xi> 0$. Although the process in (\ref{E:bm}) is not a special case of the OU process, it 
can be informally obtained from the OU process by assuming that $q=0$ and $qm=a$ in (\ref{E:OUF}). Therefore, one may say that the model, in which the volatility follows Brownian motion with drift, is an additional case of the S\&S model. We can also generate the third face of the previous model using the reflecting Brownian motion with drift as the volatility process. 
\begin{remark}\label{R:rui}
Let us denote by $p_2$ the transition density  
associated with the absolute value of Brownian motion with drift, and by $p_3$ the transition density corresponding to the reflecting Brownian motion with drift. If $a=0$, then $p_2=p_3$. This was mentioned in the introduction. For $a\neq 0$, the functions $p_2$ and $p_3$ are different. Indeed, the transition density $p_2$
is the sum of two Gaussian densities. As for $p_3$, an explicit formula is known for this density
(see formula (91) in \cite{CM}). Comparing the formulas for $p_2$ and $p_3$, we see that $p_2\neq p_3$. We refer the reader to 
\cite{GSh} and \cite{Pe} for more information the reflecting Brownian motion with drift.
\end{remark}

It is not hard to see that in the model, where the volatility follows reflecting Brownian motion with drift, we have
$\varphi_f(t)=at+\xi f(t)$, $t\in[0,T]$, and $\widehat{f}=\Gamma\varphi_f$, for all $f\in\mathbb{H}_0^1[0,T]$. 

The following lemma provides a characterization of the set $L_1$.
\begin{lemma}\label{L:char}
Let $y_0=0$. Then a function $f\in\mathbb{H}_0^1[0,T]$ belongs to the set $L_1$ defined in (\ref{E:poy}) if and only if
\begin{equation}
\dot{f}(t)\le-a\xi^{-1}
\label{E:pogh}
\end{equation}
almost everywhere on $[0,T]$ with respect to the Lebesgue measure.
\end{lemma}

\it Proof. \rm It is not hard to see that for a function $g\in\mathbb{C}_0[0,T]$, the condition $(\Gamma g)(t)=0$ holds for all $t\in[0,T]$ if and only if $g$ is a nonincreasing function on $[0,T]$.
Indeed, it follows from the definition of the Skorokhod map $\Gamma$ that the previous condition 
is equivalent to the following equality:
\begin{equation}
g(t)=\min_{0\le s\le t}(g(s)\wedge 0),\quad t\in[0,T].
\label{E:lgh}
\end{equation}
Now, it is clear that the statement formulated in the beginning of the proof can be easily derived from the equality in (\ref{E:lgh}).

Finally, by recalling the definition of the set $L_1$ and observing that the condition in
(\ref{E:pogh}) characterizes the set of functions $f\in\mathbb{H}_0^1[0,T]$, for which the function $\varphi_f(t)=at+\xi f(t)$ does not increase on $[0,T]$, we complete the proof of Lemma \ref{L:char}. 

Our next goal is to find a special representation for the rate function $\widetilde{I}_T$ under the assumption that $y_0=0$ and the reflecting Brownian motion with drift is the volatility process. 
\begin{corollary}\label{C:ler}
Consider the model, where the volatility follows the reflecting Brownian motion with drift, and let $y_0=0$. Then, the following 
formulas hold:
\begin{align*}
&\widetilde{I}_T(\mu T)=\frac{1}{2}\min\left\{\frac{a^2T}{\xi^2},
\inf_{f\in L_2}\left[\frac{\rho^2\left(\int_0^T(\Gamma\varphi_f)(s)\dot{f}(s)ds\right)^2}
{\bar{\rho}^2\int_0^T(\Gamma\varphi_f)(s)^2ds}+\int_0^T\dot{f}(s)^2ds\right]\right\}
\end{align*}
and
\begin{align*}
&\widetilde{I}_T(x)=\frac{1}{2}
\inf_{f\in L_2}\left[\frac{\left(x-\mu T-\rho\int_0^T(\Gamma\varphi_f)(s)\dot{f}(s)ds\right)^2}
{\bar{\rho}^2\int_0^T(\Gamma\varphi_f)(s)^2ds}+\int_0^T\dot{f}(s)^2ds\right],
\end{align*}
for $x\neq\mu T$.
\end{corollary}

\it Proof. \rm Corollary \ref{C:ler} follows from Corollary \ref{C:tryes}. Indeed, it only suffices to prove that
\begin{equation}
\inf_{f\in L_1}\int_0^T\dot{f}(t)^2dt=\frac{a^2T}{\xi^2}.
\label{E:lll}
\end{equation}
Using the characterization of the set $L_1$ in (\ref{E:pogh}) in Lemma \ref{L:char}, we see that for every $f\in L_1$,
$\dot{f}(t)^2\ge\frac{a^2}{\xi^2}$ a.e. on $[0,T]$. Therefore, (\ref{E:lll}) holds. 

This completes the proof of Corollary \ref{C:ler}.
\begin{remark}\label{R:bez}
The rate function in Corollary \ref{C:try} is continuous on $\mathbb{R}$, while the rate functions in Corollaries \ref{C:tryes} 
and \ref{C:ler} may be discontinuous at only one point $x=\mu T$. This can be shown 
using the same reasoning as in the proof of Lemma 4.17 in \cite{Gul1}.
\end{remark}

It remains to prove Theorem \ref{T:27}. This will be done in the next subsection.

\subsection{Proof of Theorem \ref{T:27}}\label{SS:p27}
We have already mentioned above that Theorem \ref{T:27} looks exactly like Theorem 4.2 in \cite{Gul1}. However, there are two substantial differences hidden in the formulations of those theorems. The first difference is in the structure of the mapping $f\mapsto\widehat{f}$. Recall that in the present paper, 
$$
\widehat{f}(t)=(\Gamma(G\dot{f}))(t),\quad t\in[0,T],
$$ 
where $\Gamma$ is the Skorokhod map, and $G$ is defined in Remark \ref{R:note}, while in Theorem 4.2 in \cite{Gul1}, 
$$
\widehat{f}(t)=\int_0^tK(t,s)\dot{f}(s)ds,\quad t\in[0,T],
$$ 
where $K$ is a Volterra type kernel that is Lebesgue square-integrable over $[0,T]^2$. The second difference is that Theorem \ref{T:27} uses the process 
$\varepsilon\mapsto Y^{(\varepsilon)}$ as the volatility process, while in Theorem 4.2 in \cite{Gul1}, the process 
$\varepsilon\mapsto\sqrt{\varepsilon}\widehat{B}$ is employed instead. Recall that $\widehat{B}_t=\int_0^tK(t,s)dB_s$, $t\in[0,T]$. In the present paper, the functional $\Phi$ (see (\ref{E:ref1})) and its approximation $\Phi_m$ (see (5.34) in \cite{Gul1}) are defined on the space
$\mathbb{C}_0[0,T]^2\times\mathbb{C}_{y_0}[0,T]$. Similar comparisons can be made with the proofs of LDPs in \cite{GGG}.

It follows that all the techniques employed in the proof of Theorem 4.2 in \cite{Gul1}, which do not depend on the special form of the mapping $f\mapsto\widehat{f}$, or the special structure of the volatility process, can be used in the proof of Theorem \ref{T:27}. It remains to
make a careful analysis of the proof of Theorem 4.2 in \cite{Gul1} in order to identify the statements in the proof, which depend on the above-mentioned differences, and show that those statements hold in the environment of Theorem \ref{T:27}. 

We will next prove several auxiliary lemmas. Our first goal is to estimate the distribution function of the random variable 
$$
\sup_{s\in[0,T]}Y^{(\varepsilon)}_s=||Y^{(\varepsilon)}||_{\mathbb{C}[0,T]}
$$
as $\varepsilon\rightarrow\infty$. Suppose $y> 0$, and define a subset of $\mathbb{C}[0,T]$ by
$
{\cal A}_y=\{\varphi\in\mathbb{C}:||\varphi||_{\mathbb{C}[0,T]}\ge y\}.
$
Then it is clear that the set ${\cal A}_y$ is closed in the space $\mathbb{C}[0,T]$.

The next assertion follows from the large deviation principle in Remark \ref{R:rrr}.
\begin{lemma}\label{L:lyt}
For every $y> 0$,
\begin{equation}
\limsup_{\varepsilon\rightarrow 0}\varepsilon\log\mathbb{P}(\sup_{s\in[0,T]}Y^{(\varepsilon)}_s\ge y)
\le-\inf_{\varphi\in{\cal A}_{2^{-1}y}}J(\varphi),
\label{E:09}
\end{equation}
where $J$ is defined in (\ref{E:remm}).
\end{lemma}

\it Proof. \rm It follows from (\ref{E:inad1}) and the equality 
$Y^{(\varepsilon)}=\Gamma U^{(\varepsilon)}$ that 
$$
\mathbb{P}(\sup_{s\in[0,T]}Y^{(\varepsilon)}_s\ge y)\le\mathbb{P}(\sup_{s\in[0,T]}U^{(\varepsilon)}_s\ge 2^{-1}y),\quad y> 0.
$$
Now, (\ref{E:09}), follows from the LDP in Remark \ref{R:rrr} applied to the set ${\cal A}_{2^{-1}y}$.
\begin{corollary}\label{C:lkj}
The following estimate is valid:
$$
\lim_{y\rightarrow\infty}\limsup_{\varepsilon\rightarrow 0}\varepsilon\log\mathbb{P}(\sup_{s\in[0,T]}Y^{(\varepsilon)}_s\ge y)
=-\infty.
$$
\end{corollary}

\it Proof. \rm Using Lemma \ref{L:lyt}, we see that it suffices to prove that 
\begin{equation}
\lim_{y\rightarrow\infty}\inf_{\varphi\in{\cal A}_{2^{-1}y}}J(\varphi)=\infty.
\label{E:as}
\end{equation}

We will next reason by contradiction. Suppose the equality in (\ref{E:as}) does not hold. Then, there exists a strictly increasing 
sequence $y_k> 0$, $k\ge 1$, such that $\displaystyle{\lim_{k\rightarrow\infty}y_k=\infty}$, and moreover 
\begin{equation}
\inf_{\varphi\in{\cal A}_{2^{-1}y_k}}J(\varphi)\le C,\quad k\ge 1,
\label{E:bnb}
\end{equation}
for some $C> 0$. Next, recalling the definition of $J$ in (\ref{E:remm}), we see that the estimate in (\ref{E:bnb}) can be rewritten 
as follows:
\begin{equation}
\inf_{\{\varphi:||\varphi||_{\mathbb{C}[0,T]}\ge 2^{-1}y_k\}}\inf_{\{g\in L^2[0,T]:Gg=\varphi\}}\int_0^Tg(t)^2dt\le 2C,\quad k\ge 1,
\label{E:bnbn}
\end{equation}
It follows from (\ref{E:bnbn}) that there exist two sequences $\{\varphi_k\}$ and $\{g_k\}$ such that $Gg_k=\varphi_k$,
\begin{equation}
||\varphi_k||_{\mathbb{C}[0,T]}\ge 2^{-1}y_k,\quad k\ge 1,
\label{E:lab1}
\end{equation}
and moreover, 
$
||g_k||_{\mathbb{C}[0,T]}\le C_1, 
$
for all $k\ge 1$ and some $C_1> 0$. However, by the compactness statement in Remark 
\ref{R:note}, the set $\{\varphi_k\}$ is precompact in $\mathbb{C}[0,T]$. Hence it is bounded, which contradicts (\ref{E:lab1}). 
Therefore, the equality in (\ref{E:as}) is valid.

This completes the proof of Corollary \ref{C:lkj}.

Let $\psi\in\mathbb{C}[0,T]$. The modulus of continuity of $\psi$ in $\mathbb{C}[0,T]$ is defined as follows:
$$
\widetilde{\omega}_{\delta}(\psi)=\sup_{t,s\in[0,T]:|t-s|\le\delta}|\psi(t)-\psi(s)|,\quad\delta\in[0,T].
$$
\begin{lemma}\label{L:let}
For every $y> 0$,
\begin{align*}
\lim_{\delta\rightarrow 0}\limsup_{\varepsilon\rightarrow 0}\varepsilon
\log\mathbb{P}(\widetilde{\omega}_{\delta}(Y^{(\varepsilon)})\ge y)=-\infty.
\end{align*}
\end{lemma}

\it Proof. \rm It is known that for every function $h\in\mathbb{C}[0,T]$, 
$\widetilde{\omega}_{\delta}(\Gamma\,h)\le\widetilde{\omega}_{\delta}(h)$ for all $\delta\in[0,T]$ (see, e.g., Lemma 1.1.1 (2) in \cite{P}). It follows that
\begin{align}
&\limsup_{\varepsilon\rightarrow 0}\varepsilon
\log\mathbb{P}(\widetilde{\omega}_{\delta}(Y^{(\varepsilon)})\ge y)
\le\limsup_{\varepsilon\rightarrow 0}\varepsilon
\log\mathbb{P}(\widetilde{\omega}_{\delta}(U^{(\varepsilon)})\ge y).
\label{E:loi1}
\end{align}
Set 
$$
{\cal B}_{y,\delta}=\{\varphi\in\mathbb{C}[0,T]:\widetilde{\omega}_{\delta}(\varphi)\ge y).
$$
It is not hard to prove that for every $y> 0$ and $0<\delta< T$, the set ${\cal B}_{y,\delta}$ is closed in $\mathbb{C}[0,T]$. 
Next, using the LDP in Remark \ref{R:rrr} and (\ref{E:loi1}), we obtain
\begin{align*}
&\limsup_{\varepsilon\rightarrow 0}\varepsilon
\log\mathbb{P}(\widetilde{\omega}_{\delta}(Y^{(\varepsilon)})\ge y)
\le-\inf_{\varphi\in{\cal B}_{y,\delta}}J(\varphi).
\end{align*}
It remains to prove that for every $y> 0$, 
\begin{equation}
\lim_{\delta\rightarrow 0}\inf_{\varphi\in{\cal B}_{y,\delta}}J(\varphi)=\infty.
\label{E:oo}
\end{equation}

We will next reason by contradiction. Suppose the equality in (\ref{E:oo}) does not hold. Then, it is not hard to prove, using the definition
of the rate function $J$ in (\ref{E:remm}), that there exist sequences $\delta_k> 0$, $\varphi_k\in\mathbb{C}[0,T]$, and 
$g_k\in L^2[0,T]$, $k\ge 1$, such that
the sequence $\{\delta_k\}$ is strictly decreasing, $\lim_{k\rightarrow\infty}\delta_k=0$, 
\begin{equation}
\omega_{\delta_k}(\varphi_k)\ge y,\,\,\mbox{for all}\,\,k\ge 1,
\label{E:lab2}
\end{equation}
and moreover $Gg_k=\varphi_k$ and $\int_0^Tg_k(t)^2dt\le C$, for all $k\ge 1$ and some $C> 0$. It follows from the compactness statement in 
Remark \ref{R:note} that the set $\{\varphi_k\}$ is precompact in $\mathbb{C}[0,T]$. By the Arzel\`{a}-Ascoli theorem, this set is uniformly equicontinuos. The previous statement contradicts (\ref{E:lab2}). Therefore, the equality in (\ref{E:oo}) is valid.

The proof of Lemma \ref{L:let} is thus completed.

Now, we are ready to identify the parts of the proof of Theorem 4.2 in Subsection 5.6 of \cite{Gul1}, which can not be directly transplanted into the proof of
Theorem \ref{T:27}. We will use italic font in the description of those parts below, and after every such description
include a necessary justification. \\
\\
\it The drift term $-\frac{1}{2}\varepsilon\int_0^t\sigma(s,Y^{(\varepsilon)})^2ds$ can be removed from formula (\ref{E:frrom}) not affecting the LDP (see Section 5 of \cite{G1} for a similar situation). \rm \\
\\
We can repeat the proof in Section 5 of \cite{G1} by choosing $H=\frac{1}{2}$, and replacing the estimates before (36) in \cite{G1}
by the following: For every $\delta> 0$ and the function $\eta$ defined as in Section 5 of \cite{G1}, but uniformly with respect 
to $t\in[0,T]$, 
\begin{equation}
\limsup_{\varepsilon\rightarrow 0}\varepsilon\log\mathbb{P}(\sup_{t\in[0,T]}Y^{(\varepsilon)}_t\ge\eta^{-1}(2\delta\varepsilon^{-1}T^{-1}))
=-\infty.
\label{E:reff}
\end{equation}
It is not hard to see that if we prove (\ref{E:reff}), then we can remove the drift term mentioned above exactly 
as in Section 5 of \cite{G1}. 

We will next prove the equality in (\ref{E:reff}). Set $\tau(\varepsilon)=\eta^{-1}(2\delta\varepsilon^{-1}T^{-1})$. Then $\tau$ is a strictly decreasing function on $(0,1]$. Moreover $\tau(\varepsilon)\rightarrow\infty$
as $\varepsilon\rightarrow 0$, since $\eta^{-1}(u)\rightarrow\infty$ as $u\rightarrow\infty$ (see Section 5 of \cite{G1}). Fix $\gamma> 0$ and suppose $0<\varepsilon\le\gamma$. Then $\tau(\varepsilon)\ge\tau(\gamma)$, and applying (\ref{E:09}), we see that
\begin{align*}
\limsup_{\varepsilon\rightarrow 0}\varepsilon\log\mathbb{P}(\sup_{t\in[0,T]}Y^{(\varepsilon)}_t
\ge\eta^{-1}(2\delta\varepsilon^{-1}T^{-1}))&\le
\limsup_{\varepsilon\rightarrow 0}\varepsilon\log\mathbb{P}(\sup_{t\in[0,T]}Y^{(\varepsilon)}_t\ge\tau(\gamma)) \\
&\le-\inf_{\varphi\in{\cal A}_{2^{-1}\tau(\gamma)}}J(\varphi),
\end{align*}
for all $\gamma> 0$. Finally, by taking into account (\ref{E:as}) and the fact that $\tau(\gamma)\rightarrow\infty$ as $\gamma\rightarrow 0$, 
we establish (\ref{E:reff}). It follows that the drift term mentioned above can be removed. \\
\\
\it Lemmas 5.23 and 5.24 in \cite{Gul1} hold in our setting. \rm \\
\\
Analyzing the proof of those lemmas in \cite{Gul1}, we see that the only statement in the proof that depends on the special structure of the mapping $f\mapsto\widehat{f}$ is the following: For every $\alpha> 0$,
\begin{equation}
\sup_{f\in\mathbb{H}_0^1[0,T]:||\dot{f}||_{L^2[0,T]}\le\alpha}\omega_{\frac{T}{m}}(\widehat{f})\rightarrow 0
\label{E:brr}
\end{equation}
as $m\rightarrow \infty$. The formula in (\ref{E:brr}) follows in our setting from the definition of $\widehat{f}$, the boundedness of the Skorokhod map
in $\mathbb{C}[0,T]$, the compactness statement in Remark \ref{R:note}, and the Arzel\`{a}-Ascoli theorem. \\
\\
\it Corollary 5.22 in \cite{Gul1} holds in our setting with $\sqrt{\varepsilon}\widehat{B}$ replaced by $Y^{(\varepsilon)}$. \rm \\
\\
The previous statement follows from Lemma \ref{L:let}. \\
\\
\it Lemma 5.25 in \cite{Gul1} holds in our setting with $\sqrt{\varepsilon}\widehat{B}$ replaced by $Y^{(\varepsilon)}$, and the random variables $\sigma_s^{\varepsilon,m}$ and $b_s^{\varepsilon,m}$ in formulas (5.42)-(5.44) \cite{Gul1} changed accordingly. \rm \\
\\
In the proof of the equalities similar to those in (5.51) and (5.52) in \cite{Gul1}, we use Corollary \ref{C:lkj} and Lemma \ref{L:let}, 
respectively. In our environment, the estimates similar to those for the first term on the right-hand side of (5.55) in \cite{Gul1} hold. This can be established by consulting the proof of the fact that the process in (58) of \cite{G1} is a martingale, and also the proof of (61) and 
(62) in \cite{G1}. The rest of the proof of Lemma 5.25 in \cite{Gul1} can be adapted to our environment with practically no changes. \\
\\
Finally, by taking into account what was said above, we complete the proof of Theorem \ref{T:27}.

\section{Applications}\label{S:33} 
Our first goal in the present section is to establish large deviation style formulas for binary barrier options in the small-noise regime. For Gaussian models, such result was obtained in \cite{Gul1}. Recall that the scaled asset price process $S^{(\varepsilon)}$ and the scaled log-price process $X^{(\varepsilon)}$ are defined by (\ref{E:frum}) and
(\ref{E:frrom}), respectively. It will be assumed in the present section that the drift coefficient $b$ in the model given by (\ref{E:mood})
satisfies $b(s,u)=r$, for all $(s,u)\in[0,T]\times\mathbb{R}$, where $r\ge 0$ is the interest rate. 

We will next introduce four standard forms of binary barrier options (up-and-in, up-and-out, down-and-in, down-and-out). 
Let us set the barrier at $K> 0$, and let $T> 0$ be the maturity of the option. 
\begin{definition}\label{D:dd}
Suppose the following inequality holds: $s_0< K$. \\
(i)\,The up-and-in binary barrier option pays a fixed amount $G$ of cash if the asset price process touches the barrier at some time during the life of the option. The price function of such an option in the small-noise regime is defined by
$$
V_1(\varepsilon)=Ge^{-rT}\mathbb{P}(\max_{t\in[0,T]}S_t^{(\varepsilon)}\ge K),\quad\varepsilon\in(0,T]. 
$$
(ii)\,The up-and-out binary barrier option pays a fixed amount $G$ of cash if the asset price process never touches the barrier during the life of the option. The small-noise price function in this case is given by
$$
V_2(\varepsilon)=Ge^{-rT}\mathbb{P}(\max_{t\in[0,T]}S_t^{(\varepsilon)}<K),\quad\varepsilon\in(0,T].  
$$
\end{definition}

Now, let $K< s_0$. In this case, the down-and-in and down-and-out binary options are defined similarly to the definitions 
of the up-and-in and up-and-out options given above. The price functions of the down-and-in and down-and-out options are defined by
$$
V_3(\varepsilon)=Ge^{-rT}\mathbb{P}(\min_{t\in[0,T]}S_t^{(\varepsilon)}\le K),\quad\varepsilon\in(0,T],
$$
and
$$
V_4(\varepsilon)=Ge^{-rT}\mathbb{P}(\min_{t\in[0,T]}S_t^{(\varepsilon)}> K),\quad\varepsilon\in(0,T],
$$
respectively.

Let $s_0< K$, and consider the following subsets of $\mathbb{C}_0$:
\begin{align*}
{\cal A}_T^{(1)}&=\left\{f\in\mathbb{C}_0:f(s)+x_0\ge\log K
\,\,\mbox{for some}\,\,s\in(0,T]\right\} \\
&=\left\{f\in\mathbb{C}_0:f(s)+x_0=\log K
\,\,\mbox{for some}\,\,s\in(0,T]\right\}
\end{align*}
and
\begin{align*}
&{\cal A}_T^{(2)}=\left\{f\in\mathbb{C}_0:f(s)+x_0<\log K
\,\,\mbox{for all}\,\,s\in(0,T]\right\}.
\end{align*}
Similarly, for $s_0> K$, we set 
\begin{align*}
{\cal A}_T^{(3)}&=\left\{f\in\mathbb{C}_0:f(s)+x_0\le\log K
\,\,\mbox{for some}\,\,s\in(0,T]\right\} \\
&=\left\{f\in\mathbb{C}_0:f(s)+x_0=\log K
\,\,\mbox{for some}\,\,s\in(0,T]\right\}
\end{align*}
and
$$
{\cal A}_T^{(4)}=\left\{f\in\mathbb{C}_0:f(s)+x_0>\log K
\,\,\mbox{for all}\,\,s\in(0,T]\right\}.
$$
It is not hard to see that the sets ${\cal A}_1$ and ${\cal A}_3$ are closed in the space $\mathbb{C}_0$, while
the sets ${\cal A}_2$ and ${\cal A}_4$ are open.

The next assertion provides large deviation style formulas 
for binary barrier options. 
\begin{theorem}\label{T:333}
Under the conditions in Corollary \ref{C:2} and the restrictions in the definitions of binary digital options ($s_0< K$, or $K< s_0$), 
$$
\lim_{\varepsilon\rightarrow 0}\varepsilon\log V_k(\varepsilon)=-\inf_{f\in{\cal A}_T^{(k)}}\widetilde{Q}_T(f),\quad 1\le k\le 4,
$$
where $\widetilde{Q}_T$ is the rate function given by (\ref{E:babe}).
\end{theorem}

Theorem \ref{T:333} can be established by imitating the proof of a similar result for Gaussian models 
(see the proof of Theorem 6.2 in \cite{Gul1}).
The latter proof does not use a special form of the mapping $f\mapsto\widehat{f}$, only the continuity of this mapping 
on the space $\mathbb{C}[0,T]$ is important. Note that in the proof of Theorem \ref{T:333}, we need to use the continuity of the rate function 
$\widetilde{Q}_T$ from the space $\mathbb{H}_0^1[0,T]$ into the space $\mathbb{R}$. For the Gaussian models, the previous statement
was established in \cite{Gul1}, Lemma 6.3. For the models with reflection, the proof is similar.

It is supposed in Theorem \ref{T:333} that the conditions in Corollary \ref{C:2} hold. These conditions include the assumption that the volatility function $\sigma$ is strictly positive on 
$[0,T]\times\mathbb{R}$. If follows that Theorem \ref{T:333} holds true for various models with reflection, in which the volatility function is of exponential type. However, Theorem \ref{T:333} can not be applied to the third face of the S\&S model, since the volatility function in this model, that is, the function $\sigma(x)=x$, $x\ge 0$, 
is such that $\sigma(0)=0$. For the S\&S model with reflection, a corresponding large deviation principle is that in Theorem \ref{T:27}. However, we do not know, whether the rate function $\widetilde{Q}_T$ is continuous
from the space $\mathbb{H}_0^1[0,T]$ into the space $\mathbb{R}$, under the restrictions in Theorem \ref{T:27}. 

We will next turn our attention to the following problems. Suppose the drift function in a stochastic volatility model 
with reflection (see (\ref{E:mood})) is given by $b(t,u)=r$, for all $(t,u)\in[0,T]\times\mathbb{R}^{+}$, 
where $r\ge 0$ is the interest rate.
Suppose also that the volatility function $\sigma$ satisfies the sublinear growth condition. Is the discounted asset price process 
$t\mapsto e^{-rt}S_t$ a martingale with respect to the filtration $\{{\cal F}_t\}$? Can one get a large deviation style formula for the call pricing function in such a model? We will next show that for the S\&S model with reflection, 
the answers to the previous questions are affirmative.

For the sake of convenience, let us recall that the asset price process in the S\&S model with reflection is as follows:
\begin{equation}
S_t=s_0\exp\left\{rt-\frac{1}{2}\xi^2\int_0^tY_s^2ds+\xi\int_0^tY_s
(\bar{\rho}dW_s+\rho dB_s)\right\},\,0\le t\le T,
\label{E:tre1}
\end{equation}
while the scaled version of the asset price process is given by
\begin{align}
&S_t^{(\varepsilon)}=s_0\exp\left\{rt
-\frac{1}{2}\xi^2\varepsilon\int_0^t(Y_s^{(\varepsilon)})^2ds
+\xi\sqrt{\varepsilon}
\int_0^tY_s^{(\varepsilon)}(\bar{\rho}dW_s+\rho dB_s)\right\},\,\,0\le t\le T.
\label{E:tre2}
\end{align}
In (\ref{E:tre1}) and (\ref{E:tre2}), $s_0$ is the initial condition for the asset price. Moreover,
\begin{equation}
Y_t^{(\varepsilon)}=(\Gamma U^{(\varepsilon)})(t),\quad t\in[0,T],\quad Y_0=y_0\ge 0,
\label{E:quadr2}
\end{equation}
where $\Gamma$ is the Skorokhod map, while for every $\varepsilon\in(0,1]$, 
the process $t\mapsto U^{(\varepsilon)}_t$ is the solution to the following stochastic integral equation:
\begin{equation}
U_t^{(\varepsilon)}=y_0+q\int_0^t(m-(\Gamma U^{(\varepsilon)})(s))ds+\sqrt{\varepsilon}
\xi B_t,\quad t\in[0,T]
\label{E:quad30}
\end{equation}
(see (\ref{E:quad3})). The process $Y$ appearing in (\ref{E:tre1}) is given by $Y_t=(\Gamma U^{(1)})(t)$, $t\in[0,T]$.

For the S\&S model with reflection, the call pricing function in the small-noise regime is defined by
$$
C^{(\varepsilon)}(T,K)=e^{-rT}\mathbb{E}\left[(S^{(\varepsilon)}_T-K)^{+}\right],\quad\varepsilon\in(0,1],
$$
where $T> 0$ is the maturity of the option, $K> 0$ is the strike price, and for every $u\in\mathbb{R}$, $u^{+}=\max(u,0)$. 
We assume that
$T$ and $K$ are fixed, and study the asymptotic behavior of the call pricing function when $\varepsilon\rightarrow 0$.  
\begin{theorem}\label{T:cju}
The following statements hold true for the S\&S model with reflection: \\
(i)\,The discounted asset price process $t\mapsto e^{-rt}S_t$ is a martingale with respect to the filtration
$\{{\cal F}_t\}$. \\
(ii)\,Suppose $K> 0$ and $y_0> 0$. Then
\begin{equation}
\lim_{\varepsilon\rightarrow 0}\varepsilon\log C^{(\varepsilon)}(T,K)=-\inf_{x:x\ge\log K-x_0}\widetilde{I}_T(x),
\label{E:when}
\end{equation}
where $\widetilde{I}_T$ is the rate function in Corollary \ref{C:try}. \\
(iii)\,Suppose $K> 0$ and $y_0=0$. Suppose also that the call option is out-of-the-money, that is, $K>s_0e^{rT}$. 
Then, the equality in (\ref{E:when}) holds with the rate function 
$\widetilde{I}_T$ given in Corollary \ref{C:tryes}.
\end{theorem}
\begin{remark}\label{R:88}
It follows from part (i) of Theorem \ref{T:cju} that $\mathbb{P}$ is a risk-neutral measure for the S\&S model with reflection.
Parts (ii) and (iii) provide large deviation style formulas for the call pricing function in the small-noise regime.
\end{remark}

\it Proof of Theorem \ref{T:cju}. \rm Set 
$\displaystyle{B^{*}_t=\max_{0\le s\le t}|B_s|}$, $t\in[0,T]$.
We will need the following lemma.
\begin{lemma}\label{L:gron}
For every $\varepsilon\in(0,1]$ and $t\in[0,T]$, the following estimate holds $\mathbb{P}$-a.s. on $\Omega$:
\begin{equation}
\max_{0\le s\le t}Y_t^{(\varepsilon)}\le 2e^{2qt}y_0+m(e^{2qt}-1)+2\sqrt{\varepsilon}\xi e^{2qt}B_t^{*}.
\label{E:asu}
\end{equation}
\end{lemma}

\it Proof. \rm For all $\varepsilon\in(0,1]$ and $t\in[0,T]$, denote 
$\displaystyle{Z_t^{(\varepsilon)}=\max_{0\le s\le t}|U_s^{(\varepsilon)}|}$.
Then, using (\ref{E:inad1}) and (\ref{E:quadr2}), we get
\begin{equation}
Y_t^{(\varepsilon)}\le 2Z_t^{(\varepsilon)},\quad\varepsilon\in(0,1],\quad t\in[0,T].
\label{E:jkl}
\end{equation}
It follows from (\ref{E:quad30}) that
$$
|U_t^{(\varepsilon)}|\le y_0+qmt+2q\int_0^tZ_s^{(\varepsilon)}ds+\sqrt{\varepsilon}\xi|B_t|,
$$
and therefore
\begin{equation}
Z_t^{(\varepsilon)}\le y_0+qmt+\sqrt{\varepsilon}\xi B_t^{*}+2q\int_0^tZ_s^{(\varepsilon)}ds,
\label{E:gro}
\end{equation}
for all $t\in[0,T]$.

Our next goal is to apply Gronwall's inequality to (\ref{E:gro}). We will use the following version of Gronwall's lemma
(see \cite{CL}, p.37). Let $\varphi$, $\psi$, and $\chi$ be real-valued continuous functions on the interval $[a,b]$. Suppose
$\chi(t)> 0$, and 
$$
\varphi(t)\le\psi(t)+\int_a^t\chi(s)\varphi(s)ds,
$$
for all $t\in[a,b]$. Then
$$
\varphi(t)\le\psi(t)+\int_a^t\chi(s)\psi(s)\exp\left\{\int_s^t\chi(u)du\right\}ds,
$$
for all $t\in[a,b]$.

Let $[a,b]=[0,T]$, $\varphi(t)=Z_t^{(\varepsilon)}$, $\chi(t)=2q$, and $\psi(t)=y_0+qmt+\sqrt{\varepsilon}\xi B_t^{*}$.
Next, applying Gronwall's lemma to (\ref{E:gro}), we obtain the following estimate:
\begin{align*}
&Z_t^{(\varepsilon)}\le y_0+qmt+\sqrt{\varepsilon}\xi B_t^{*}+2q\int_0^t(y_0+qms+\sqrt{\varepsilon}\xi B_s^{*})\exp\{2q(t-s)\}ds
\\
&=y_0+qmt+\sqrt{\varepsilon}\xi B_t^{*}+(e^{2qt}-1)y_0+2q^2m\int_0^ts\exp\{2q(t-s)\}ds+\xi\sqrt{\varepsilon}(e^{2qt}-1)B_t^{*}.
\end{align*}
Using the integration by parts formula in the integral on the previous line and simplifying the resulting expression, we see that
$$
Z_t^{(\varepsilon)}\le e^{2qt}y_0+m\frac{e^{2qt}-1}{2}+\xi\sqrt{\varepsilon}e^{2qt}B_t^{*}.
$$
Finally, by taking into account (\ref{E:jkl}), we get
$$
Y_t^{(\varepsilon)}\le 2e^{2qt}y_0+m(e^{2qt}-1)+2\xi\sqrt{\varepsilon}e^{2qt}B_t^{*}.
$$
Now, it is clear that (\ref{E:asu}) follows from the previous estimate.

The proof of Lemma \ref{L:gron} is thus completed.
\begin{corollary}\label{C:fgh}
There exists $\alpha> 0$ such that 
\begin{equation}
\mathbb{E}\left[\exp\left\{\alpha\sup_{\varepsilon\in(0,1]}\max_{0\le t\le T}(Y_t^{(\varepsilon)})^2\right\}\right]<\infty.
\label{E:lpt}
\end{equation}
\end{corollary}

\it Proof. \rm Using the estimate in (\ref{E:asu}), we see that in order to prove (\ref{E:lpt}), it suffices to show that
there exists $\beta> 0$ such that
$$
\mathbb{E}\left[\exp\left\{\beta(B_T^{*})^2\right\}\right]<\infty.
$$

The following estimate is known (see, e.g., \cite{H}, p. 31):
\begin{equation}
\mathbb{P}(B^{*}_T> y)\le\frac{4}{\sqrt{2\pi T}}\int_y^{\infty}\exp\left\{-\frac{z^2}{2T}\right\}dz,
\label{E:ho}
\end{equation}
for all $y> 0$. The inequality in (\ref{E:ho}) can be established as follows. It is not hard to see that 
$$
\{B^{*}_T> y\}\subset\{\max_{0\le t\le T}B_t> y\}\cup\{\min_{0\le t\le T}B_t<-y\}=\{\max_{0\le t\le T}B_t> y\}
\cup\{\max_{0\le t\le T}(-B_t)> y\}.
$$
Since the process $-B$ is also a Brownian motion, the previous inclusion and the reflection principle imply (\ref{E:ho}).
It follows from (\ref{E:ho}) that
\begin{equation}
\mathbb{P}(B^{*}_T> y)=O\left(\exp\left\{-\frac{y^2}{2T}\right\}\right),
\label{E:cvg}
\end{equation}
as $y\rightarrow\infty$. 

Choose $\beta<\frac{1}{2T}$. Next, using (\ref{E:cvg}) and the integration by parts formula, we obtain
\begin{align*}
&\mathbb{E}\left[\exp\left\{\beta(B_T^{*})^2\right\}\right]=-\int_0^{\infty}\exp\{\beta y^2\}d\mathbb{P}(B^{*}_T> y) \\
&=1+2\beta\int_0^{\infty}y\mathbb{P}(B^{*}_T> y)\exp\{\beta y^2\}dy<\infty.
\end{align*}

This completes the proof of Corollary \ref{C:fgh}.

We will next return to the proof of Theorem \ref{T:cju}. According to (\ref{E:tre1}) the discounted asset price process 
has the following form:
\begin{equation}
e^{-rt}S_t=s_0\exp\left\{-\frac{1}{2}\xi^2\int_0^tY_s^2ds+\xi\int_0^tY_s
(\bar{\rho}dW_s+\rho dB_s)\right\},\,0\le t\le T.
\label{E:sto}
\end{equation}
It follows from Corollary \ref{C:fgh} that for some $\alpha> 0$,
$$
\mathbb{E}\left[\exp\left\{\alpha\max_{0\le t\le T}(Y_t)^2\right\}\right]<\infty.
$$
The previous inequality implies that the stochastic exponential on the right-hand side of (\ref{E:sto}) is a martingale (see, e.g.,
\cite{G}, Corollary 2.11).

This completes the proof of part (i) of Theorem \ref{T:cju}.

We will next turn our attention to parts (ii) and (iii) of Theorem \ref{T:cju}. The large deviation principles in Corollaries \ref{C:try}
and \ref{C:tryes} will be used in the proofs. We will only sketch these proofs since there exist well-known 
methods allowing to derive asymptotic formulas for call pricing functions from large deviation principles. For such derivations, 
we refer the reader to \cite{P}, the proof on p. 36; \cite{FZ}, Corollary 4.13; \cite{G2}, Section 7, pp. 1131-1133; or \cite{Gul1}, 
part (i) of Theorem 5.2.

The out-of-the-money condition $K> s_0e^{rT}$ appears in part (iii) of Theorem \ref{T:cju} because we do not know whether the rate function
$\widetilde{I}_T$ defined in Corollary \ref{C:tryes} is continuous at $x=rT$ for $y_0=0$ (see Remark \ref{R:bez}). However, the
rate function in Corollary \ref{C:try} is continuous everywhere on $\mathbb{R}$, and no extra restrictions are needed here. It follows from the previous remark that if $y_0> 0$, then the set $[\log K-x_0,\infty)$
is a set of continuity for the rate function $\widetilde{I}_T$ for all $K> 0$, while if $y_0=0$, then the same set is 
a set of continuity for $\widetilde{I}_T$ for all $K> s_0e^{rT}$. 
Note that the condition $K> s_0e^{rT}$ implies the inequality $rT<\log K-x_0$. 

The first step in the proof of parts (ii) and (iii) is to establish a large deviation style formula for the binary call option. The pricing function for such an option is defined by 
$$
c^{(\varepsilon)}(T,K)=e^{-rT}\mathbb{P}(S_T^{(\varepsilon)}\ge K)=e^{-rT}\mathbb{P}\left(X_T^{(\varepsilon)}-x_0\ge(\log K)-x_0\right),
$$ 
where $K> 0$ is the strike price and $T> 0$ is the maturity of the option. The following formula holds for all $K> 0$ provided that
$y_0> 0$:
\begin{equation}
\lim_{\varepsilon\rightarrow 0}\varepsilon\log c^{(\varepsilon)}(T,K)=-\inf_{x:x\ge\log K-x_0}\widetilde{I}_T(x).
\label{E:outo}
\end{equation}
In addition, if $y_0=0$, then the formula in (\ref{E:outo}) is valid when the option is out-of-the money. The above-mentioned formulas  follow from the large deviation principles in Corollaries \ref{C:try}
and \ref{C:tryes}. The proof also uses the continuity properties of the function $\widetilde{I}_T$ (see the discussion above).

It remains to derive the formulas in parts (ii) and (iii) of Theorem \ref{T:cju} from the formula in (\ref{E:outo}).
The lower large deviation estimate for the call pricing function can be obtained from (\ref{E:outo}) and the continuity properties of the rate function.
In addition, to prove the upper estimate we use H\"{o}lder's inequality, (\ref{E:outo}), the continuity properties of the rate function, 
part (i) of Theorem \ref{T:cju}, and Corollary \ref{C:fgh}. More details can be found in \cite{P}, \cite{FZ}, or (\cite{Gul1}). 

This completes the proof of Theorem \ref{T:cju}.

\end{document}